\documentstyle[pre,aps,epsf]{revtex}
\newcommand{\cO}{{\cal{O}}}
\newcommand{\BEQ}{\begin{equation}}
\newcommand{\EEQ}{\end{equation}}
\newcommand{\BEA}{\begin{eqnarray}}
\newcommand{\EEA}{\end{eqnarray}}
\renewcommand{\H}{{\cal {H}}}

\newcommand{\m}{{\tilde m}}

\renewcommand{\S}{S_{\em ep}}
\newcommand{\p}{\partial}

\newcommand{\I}{{\cal {I}}}
\newcommand{\nn}{\nonumber }
\newcommand{\Kt}{{\tilde K}}
\newcommand{\Ht}{{\tilde H}}

\newcommand{\bam}{{\overline m}}
\newcommand{\baG}{{\overline {\Gamma}}}
\newcommand{\baTe}{{\overline {T}_e}}
\newcommand{\bamu}{{\overline {\mu}}_2}
\newcommand{\baw}{{\overline {w}}}
\newcommand{\tinf}{{\mbox{\hspace*{-0.2 mm}\tiny $\infty$}}}
\newcommand{\tp}{{\mbox{\hspace*{-0.2 mm}\tiny $+$}}}
\def\dbarrm {{\mathchar'26\mkern-11mu{\rm d}}}                         
                               
\renewcommand{\thesection}{\arabic{section}}

\renewcommand{\theequation}{\thesection\arabic{equation}}

\begin{document}
\title{Effective temperatures in an exactly solvable
 model for a fragile  glass}
\author{Luca Leuzzi   and    Theo~M.~Nieuwenhuizen\\
Universiteit van Amsterdam
\\ Valckenierstraat 65, 1018 XE Amsterdam, The Netherlands}
\date{  printout: \today}
\maketitle
\begin{abstract}
A  model glass is considered with one type of fast ($\beta$-type)
of processes, and one type of slow processes ($\alpha$-type).
On time-scales where the fast ones are in equilibrium, the slow ones
have a dynamics that resembles the one of facilitated spin models.
The main features are the occurrence of a Kauzmann transition, 
a Vogel-Fulcher-Tammann-Hesse behaviour for the relaxation time, 
an Adam-Gibbs relation between relaxation time and configurational
entropy,  and an aging regime.
The model is such that its statics is simple and its 
(Monte-Carlo-type) dynamics is exactly solvable.
The dynamics has been studied both on the approach 
to the Kauzmann transition and below it.
In certain parameter regimes it is so
slow that it sets out a quasi-equilibrium at a
time dependent {\em effective temperature}.
Correlation and Response functions are also computed, as well as
 the out of equilibrium
Fluctuation-Dissipation Relation, showing the uniqueness of
the effective temperature, thus giving support to the rephrasing
of the problem within the framework 
of out of equilibrium thermodynamics.

\end{abstract}

\renewcommand{\thesection}{\arabic{section}}
\section{Introduction}
\setcounter{equation}{0}\setcounter{figure}{0}
\renewcommand{\thesection}{\arabic{section}.}
\label{sec1}

A glass can be viewed as a liquid in which a huge slowing down  of the
 diffusive motion of the particles has destroyed its ability to flow
on experimental time-scales.
The slowing down can be expressed through the relaxation time, i. e.  
the characteristic time needed to have one inter-particle diffusion process
of a particle while it is rattling between its neighbour
particles, that  form a  cage around it. 
This relaxation time is proportional to viscosity.
 Cooling down from the liquid phase,  at some point 
 the system falls out of equilibrium: the slow
liquid degrees of freedom are no more accessible and
the relaxation time and the   viscosity of the under-cooled melt grow
suddenly  by several orders of magnitude. 
The temperature at which this happens is defined as the 
glass transition temperature $T_g$ \cite{ANGELL}. At $T_g$
 the heat capacity  decreases in a clear way
 going from liquid to glassy phase
 and  also on reheating an abrupt, but different change shows up.
(Some universal behavior in the cooling-heating process was pointed
 out by one of us~\cite{NPRL98,N00}.) Moreover, discontinuities of 
this kind occur also in the compressibility and the thermal expansivity.
This looks similar to a continuous phase transition, even though the 
analogy is not perfect, because of the smeared nature of the 
discontinuities and because the smaller specific heat value occurs
below the glass transition, rather than above, as would normally occur
in mean field phase transitions.

The above described transition is not a true thermodynamic phase 
transition, but it is strictly kinetic in origin: it takes place when 
the relaxation time becomes longer than the observation time
and marks the transition from ergodic to non-ergodic behaviour.
In general the location of this transition, the empirical 
glass transition temperature $T_g$, depends on the cooling rate,
more precisely, on the cooling scheme.
The absolute glass transition temperature
is defined as the temperature where the viscosity equals $10^{13}$
Poise, and the equilibration time of the order of days.
It is related to the slowest possible experiments one can 
realistically do. Cooling at higher rates there is a glass transition
at a somewhat larger dynamical glass transition temperature.

The very slow relaxation of non-ergodic systems
 evolving towards equilibrium structures
 on  time scales longer than the characteristic time scales of
the experiments,  depending on the history of the system
(e. g. on the phase space region  in which the initial
 conditions are chosen and on the cooling rate)
is the so called regime of aging dynamics \cite{BBM,BCKM}.

Experimental data for the viscosity pattern of glass forming liquids
are often fit to a Vogel-Fulcher-Tammann-Hesse (VFTH)
behavior\cite{VFTH1,VFTH2,VFTH3}: $\tau_{\rm eq}$ $\sim$
$\exp [A^\gamma/(T-T_0)^{\gamma}]$, where 
the fitting parameter $T_0$  depends on the 
material and the range of temperatures in which the fit is performed.
The exponent $\gamma$ is usually set equal to $1$, 
and an argument for this choice 
was given by  Adam and Gibbs\cite{AG}. However also
 exponents $\gamma \neq 1$ are compatible with data,
 merely affecting the width of the fitting interval.  An analytic  
approach by Parisi\cite{ParisiVF}, on the other hand, using replica trick 
and field theory, gives  $\gamma=2$ in three dimensions.
Here we shall consider $\gamma$ as a model parameter, that can be
chosen below, equal to or above unity, and 
investigate the  aspects of this standard picture.

Kauzmann ~\cite{KAUZ} pointed at the paradox that the
difference between the liquid entropy and the crystal entropy 
(i.e. the entropy of the most organized state for the system) 
if naively extrapolated to zero temperature would
 become negative at some point. To circumvent this unphysical result 
he proposed the occurrence of a thermodynamic phase transition at 
the temperature (commonly denoted by $T_K$) 
where this entropy difference vanishes.
Such a thermodynamic transition would be characterized by a
discontinuity of the specific heat and by the exponential divergence of the
relaxation time  (VFTH  for fragile glasses or Arrhenius for strong ones). 
Connected to this last feature is the usual assumption that 
the fitting parameter $T_0$ of the VFTH law 
coincides with the Kauzmann temperature $T_K$.
We note that at this phase transition, the divergence of the relaxation time 
 is  not algebraic in temperature, as happens in  ordinary continuous 
phase transitions, but it is exponential and no
   susceptibility diverges at the critical point.

The residual entropy given by the difference of the entropy of the 
undercooled liquid and the entropy of the vibrational modes of the
 crystal that could in principle be formed, is usually called 
complexity or configurational entropy. 
According to standard knowledge, the Kauzmann transition should
be characterized by a vanishing, or minimal, configurational entropy. 
This prediction is very difficult to test experimentally, 
since the relaxation time is too long. The existence of a Kauzmann 
transition was nevertheless recently supported  both by analytical and 
numerical  results  \cite{MP,CPV1,CPV2}. The configurational entropy
 is the entropy determined by the number of states that the system at 
 temperature $T_K<T<T_g$  can visit.

At a given  dynamic critical temperature $T_D$, generally greater than 
$T_g$, the separation of the time scales of slow ($\alpha$)
and fast ($\beta$) processes starts to increase more rapidly than 
at higher temperature. 
Referring to  the phase space 
we can say that structures get organized at two levels:
some minima of the free energy are separated by very small barriers 
and between them $\beta$ processes take place; 
groups of those minima are contained in bigger basins
separated by barriers requiring a greater free energy 
variation to be crossed.
 To make the system go from the configuration
in one of these basins to another configuration in another basin, i. e.
to have an $\alpha$ process, a longer time is needed.
 The time scale on which these processes are 
happening are, however, at $T_D$ and below (but above $T_g$),
 still very short in comparison with
the observation time. In a cooling experiment the system
 is thus still in thermodynamic equilibrium. 
Going on with cooling, the deepness of the local and global minima, 
appearing in the  thermodynamic potential and corresponding to
 different metastable  and stable states, grows: 
barriers between them become higher and higher
 until, at the glass transition temperature $T_g$,
 some states become impossible to reach during the time-scale
 we set for our system, i. e. the experimental time. 
The configurational entropy is the observable that counts the relevant
states.
As temperature decreases further the configurational entropy
starts to decrease because the states available for the system 
are less and less. 
The Kauzmann temperature is reached when the system
 is stuck in one  state and cannot move to any other, because, even
 asymptotically and even for short range systems where activated processes
were present for $T_K<T<T_g$, the free energy barriers become infinite.

The configurational entropy density $s_{c}$ is usually connected 
directly to the relaxation time through the Adam-Gibbs relation 
\cite{AG}: $ \tau_{eq}\sim \exp\left(1/s_{c}\right)$.

To recapitulate we have been referring basically
to the following different regimes for a glass former.
\begin{itemize}
\item For $T>T_D$ the system  is in  a disordered phase. Diffusion processes
have a very short relaxation time. At very high temperature
the free energy describing 
the system has only one global equilibrium minimum and cooling towards
the $T_D$ temperature small local minima show up.

\item Around $T_D$ a dynamic transition takes place. The phase is still
disordered but
the free energy increases its roughness and some local minima become deeper:
$\alpha$-$\beta$ bifurcation is qualitatively enhanced.
In a simple mode coupling theory\cite{GO} this is the 
temperature $T_{mc}$ at which a static transition is predicted with 
an algebraically diverging relaxation time.
In the p-spin spin glass
model \cite{CHS,KT}  this corresponds to the dynamic critical temperature
$T_D$ at which the system goes to metastable states of energy higher
than the minimum energy. For $T_g<T<T_D$ the dynamics of $\alpha$ processes
has a huge slowing down but the temperature is high
enough to reach equilibrium on the experimental time scales.

\item Around  $T_g$, that depends on the cooling rate,
 another transition takes place. Many other local minima appear and 
the free energy roughness is such that the deepest local minima,
corresponding to metastable states, become ergodically separated on the time 
scales of the experiment.
For $T_K<T<T_g$ the system has a very slow aging dynamics between the 
metastable states, proceeding by activated processes.

\item At $T=T_K$  a thermodynamic phase transition shows up,
 with exponentially diverging relaxation time. 
The free energy barriers between deep local 
minima increase to infinite and the system gets stuck in one single minimum
forever. Ergodicity is broken at any time scale.
In the p-spin model this corresponds to the temperature at which 
the replica symmetry is broken \cite{CS}.
The Kauzmann temperature $T_K$ 
is usually assumed to coincide with the fitting
parameter $T_0$ of the VFTH law.
Below the Kauzmann temperature the system evolves only  through the 
configurations belonging to the ergodic component of the phase space
where the dynamics  brought it during the cooling.

\item In a cooling experiment that goes below $T_K$, and in a
quenching experiments to a temperature below $T_K$, the aging dynamics
 visits only states with a free energy that is  higher  than the one
in the static limit. Dynamics behaves as if occurring at a higher
 temperature.

\end{itemize}

The exponential divergence of  time scale in glasses 
(opposed to  the algebraic divergence in  standard
continuous phase transitions)
might induce an asymptotic decoupling of the time-decades.
The reasonable assumption can be made that, in a glassy
system that has aged a long time $t$,  all processes
with equilibration time much less than $t$ are in equilibrium (the
$\beta$ processes), while those evolving on 
 time scales much larger than $t$ are still
quenched, leaving the processes with time scale of order $t$ 
(i. e. the $\alpha$ processes), as the only interesting ones. 
Indeed this assumption was already tested successfully in  models similar to, 
but even more idealized than, the one we are going to discuss in
the present paper. Those models showed a glassy regime  with an 
Arrhenius law (rather than VFTH),
like a harmonic oscillator model \cite{BPR,N00} and a
spherical spin model \cite{NPRL98,N00}.
The asymptotic decoupling of time scales that is the input for the
present set of models could be the basic ground for 
a generalization of equilibrium thermodynamics to systems 
out of equilibrium \cite{N00}.
That approach involves systems where one extra variable is
needed to describe the non-equilibrium physics, namely the
{\em effective temperature}. One of our aims will be to test this picture
in an exactly solvable model glass; 
 we shall see that there are domains where it does apply
(namely when the VFTH exponent $\gamma$ exceeds unity) and where it
does not apply (namely when $\gamma\le 1$). 
In this last case two
extra variables will be needed making compulsory the introduction of an {\em 
effective field} besides the effective temperature.

In the present paper we are going to investigate  an exactly
 solvable  model glass that shows all of the features that we recalled above
for the much more complicated real glasses.
The model is introduced in section  \ref{sec2}. It is built 
 by processes evolving on two different, well separated time scales,
 representing the $\alpha$ and $\beta$ processes taking place in real
glassy materials.
In section \ref{sec3} we introduce the  dynamics that we 
apply to the model and we show the  dynamic behaviour
in the aging regime. We can implement the dynamics even below the
Kauzmann temperature, thus getting insight in a regime where few
 analytic results are known. Even though the physics of our model is
 simple, we shall find general aspects of the results
by formulating them in the thermodynamic language. 
The  phrasing of the dynamic properties in terms of a generalized 
 out-of-equilibrium thermodynamic frame is carried out
 in section \ref{sec4}, where we introduce effective parameters to 
take into account the history of the system.
In section \ref{sec5} we study two-time observables, such as
correlation functions and response functions,
and we look at the fluctuation dissipation ratio
out of equilibrium \cite{CK93}.


\renewcommand{\thesection}{\arabic{section}}
\section{Model}
\setcounter{equation}{0}\setcounter{figure}{0}
\renewcommand{\thesection}{\arabic{section}.}
\label{sec2}



The model we study, that was firstly introduced in \cite{NCM}, 
is described by the following local Hamiltonian

\BEQ
{\cal{H}}[\{x_i\},\{S_i\}]=\frac{1}{2} K
\sum_{i=1}^{N}x_i^2
-H\sum_{i=1}^{N}x_i
-J\sum_{i=1}^{N}x_iS_i
-L\sum_{i=1}^{N}S_i
\label{sec2:Hmodel}
\EEQ

\noindent where $N$ is the size of the system and 
$\{x_i\}$ and $\{S_i\}$ are continuous variables, 
the last satisfying a spherical constraint: $\sum_i S_i^2=N$.
We will call them from now on  harmonic oscillators
and spherical spins, respectively.
$K$ is the Hooke elastic constant, $H$ is an external field acting on
the harmonic oscillators, $J$ is the coupling constant between $\{x_i\}$ 
and $\{S_i\}$ and $L$ is the external field acting on the spherical spins.
As we will see in this paper the  
simple local form of (\ref{sec2:Hmodel})  allows us to introduce
 an analitycally solvable dynamics with glassy behaviour.

In our simple model we  introduce by hand a separation
of time scales where the spins represent the fast modes and  the
harmonic oscillators the slow ones.
Separation of time scales is one of the most important and most general 
characteristics  that glasses are supposed to hold.
Indeed, we  assume that the $\{S_i\}$  evolve with time on a much shorter 
time scale than that of the harmonic oscillators.
From the point of view of the spins the $\{x_i\}$ are quenched random
variables and the combination $Jx_i$ can be seen as a random field exerted 
 on spin $i$.
On the other hand, from the point of view of the motion of the $\{x_i\}$ the spins are just
a noise. 
To describe the long time regime of the $\{x_i\}$ system
 we can average over this noise by performing the computation 
of the $\{S_i\}$ partition function,
yielding an effective Hamiltonian depending only on the $\{x_i\}$,
that will determine the dynamics of these variables.

Summing out fast variables is a standard technique in physics. For
instance, in any Landau-Ginzburg-Wilson theory there occur
coefficients, of which the temperature dependence arises from summing
out fast processes. We now do the same in our model.

We perform the spin integration in the partition function using the 
saddle point approximation for large $N$ and we get:
\BEA
Z_S(\{x_i\})&=&
\int\left(\prod_{i=1}^{N} dS_i\right)
 \exp\left\{-\beta{\cal H}\left[\{x_i\},\{S_i\}\right]\right\}\,\,
\delta\left(\sum_{i=1}^{N}S_i^2-N\right)
\\ \nn
&\simeq&
\exp\left[-\beta N\left(
\frac{K}{2}  m_2  - H  m_1  - w +
 \frac{T}{2}\log\frac{w+\frac{T}{2}}{T} \right)\right]
\EEA
With $\beta=1/T$ and where we introduced the short-hands
\BEQ
m_1\equiv\frac{1}{N}\sum_{i=1}^Nx_i, \hspace*{2 cm}
m_2\equiv\frac{1}{N}\sum_{i=1}^Nx_i^2 \ .
\label{sec2:abbrev_m}
\EEQ
and 
\BEQ
w\equiv\sqrt{J^2 m_2+2JLm_1+ L^2+ \frac{T^2}{4}} \ .
\EEQ        
We can define the effective Hamiltonian
${\cal{H}}_{\rm eff}(\{x_i\})\equiv -T\log Z_S(\{x_i\})$, obtaining
\BEQ               
{\cal{H}}_{\rm eff}(\{x_i\})=
\frac{K}{2}  m_2 N - H  m_1 N - w N+
 \frac{TN}{2}\log\frac{w+\frac{T}{2}}{T}
\label{sec2:Heff}
\EEQ                                                             
This can also be written  in terms of the internal energy $U(\{x_i\})$
and of the entropy 
of the equilibrium processes (i.e. the spins) $S_{\rm ep}(\{x_i\})$:
\BEA
{\cal{H}}_{\rm eff}(\{x_i\})&=&U(\{x_i\})-T S_{\rm ep}(\{x_i\})
\label{def:Heff}\\
U(\{x_i\})&=&N\left[ \frac{K}{2}  m_2  - H m_1  - w + \frac{T}{2}\right]
\label{def:U}\\
S_{\rm ep}(\{x_i\})&=&\frac{N}{2} \left[1-\log \frac{w+T/2}{T}\right]
\label{def:Sep}
\EEA
and it can indeed be verified that $U$ is the Hamiltonian 
averaged over the spins and that $S_{\rm ep}$ is the entropy of the spins. 

Another fundamental ingredient for the  model is the introduction of  
a constraint on the phase space to avoid the existence of the single 
global minimum $\{ x_i=0,  \forall i\}$, thus
implementing a large degeneracy of the allowable lowest states. 
The constraint is taken on the $\{x_i\}$,
 thus concerning the long time regime. It reads:
\BEQ
m_2-m_1^2\geq m_0 
\label{CONSTRAINT}
\EEQ
where $m_0$ is a fixed but arbitrary, strictly positive constant.
The now obtained model glass has no crystalline state.
This constraint applied to the harmonic oscillators dynamics is
a way to reproduce the behaviour of good glass formers, 
i. e. substances for which nucleation of the crystal phase
is especially unlikely even at very slow cooling rates
(e. g. network formers $B_2O_3$ and  $Si O_2$, molecule organics
such as glycerol and  atactic polystyrene and  different multicomponent
 liquid mixtures). 
These are  substances for which there are 
 non-crystalline packing modes for the particles composing them 
that have intrinsically low energy, thus favouring these disordered
configurations. In general the crystal state still exists, at
lower energy, but the probability of nucleating a crystal 
instead of a glass is  negligible.
 In specific cases 
(binary solutions) the glassy state can even be lower in energy 
than the crystalline one and be thermodynamically stable with 
respect to any crystal configuration \cite{KA}.

As we will explain in detail in the next section, we impose 
a dynamics which satisfies this constraint and couples 
the otherwise non-interacting $\{x_i\}$ in a dynamic way.

To shorten the notation 
for later purposes we define here the modified ``spring constant'' 
${\tilde{K}}$ and ``external field'' ${\tilde{H}}$:
\BEQ
{\tilde{K}}=K-\frac{J^2}{w+T/2},\hspace*{2 cm}
{\tilde{H}}=H+\frac{JL}{w+T/2}
\label{def:tilde}
\EEQ
We stress  that ${\tilde{K}}$ and ${\tilde{H}}$ are actually functions of
the $\{x_i\}$ themselves (through $m_1$ and $m_2$ that occur  in $w$).
We also define the constant
\BEQ
D\equiv HJ+KL 
\label{D}
\EEQ
Recalling  the definitions
 (\ref{def:tilde})
it is useful  to note 
that 
\BEQ
{\tilde{H}}J+{\tilde{K}}L=HJ+KL=D.
\EEQ
\subsection{Statics}

The partition function of the whole system at equilibrium is:
\BEA
Z(T)&&=\int {\cal{D}}x {\cal{D}} S \exp\left[-\beta{\cal{H}}(\{x_i\},\{S_i\})\right] \delta\left(\sum_i x_i^2-N\right)=
\label{Z_static}
\\
\nn
&&=
\int d m_1 d m_2 \exp\left\{-\beta N\left[\frac{K}{2} m_2 -H m_1-w+\frac{T}{2}\log\left(\frac{w+T/2}{T}\right)-\frac{T}{2}\left(1+\log(m_2-m_1^2)\right)\right]\right\}
\EEA
The new object that appears in the exponent is the 
configurational entropy
\BEQ
{\cal{I}}\equiv\frac{N}{2}\left(1+\log(m_2-m_1^2)\right)
\label{def:I}
\EEQ
 which will be widely considered in section \ref{sec4}. 
It comes  from the 
Jacobian of the transformation of variables
${\cal{D}}x \to dm_1dm_2$, $\exp{\I}$, 
(see (\ref{sec2:abbrev_m})).
We can compute the large $N$ limit of this partition using once again
the saddle point 
approximation.
The saddle point equations are found minimizing 
the function
\BEQ
\frac{\beta}{N}F(T,m_1,m_2)\equiv
\frac{1}{T}\left(\frac{K}{2} m_2 -H m_1-w\right)
+\frac{1}{2}\left[+\log\frac{w+T/2}{T}-1-\log(m_2-m_1^2)\right]
\EEQ
Denoting the saddle point values of $m_1$ and $m_2$ as $\bam_1$ and 
$\bam_2$ the equations  are:
\BEA
\bam_1&&=\frac{\Ht(\bam_1,\bam_2)}
{\Kt(\bam_1,\bam_2)}
\label{SPeq1}
\\
\bam_2&&=
\bam_1^2+
\frac{T}
{\Kt(\bam_1,\bam_2)}\label{SPeq2}
\EEA
The form of the solutions $\bam_1(T)$, $\bam_2(T)$
 is quite complicated because each of these
equations is actually a fourth order equation, but they can 
be explicitely computed.
In terms of the equilibrium values $\bam_k$ we find the following expression for the equilibrium free energy density:
\BEA
F(T,\bam_1(T),\bam_2(T))&&=N\left[\frac{K}{2} \bam_2 -H \bam_1-
w\left(\bam_1,\bam_2\right)\right]
+\frac{TN}{2}\left[\log
\frac{w\left(\bam_1,\bam_2\right)+T/2}{T}
-1-\log(\bam_2-(\bam_1)^2)\right]
\\
&&=\hspace*{ 1 cm} U(T,\bam_1,\bam_2)\hspace*{ 1.15 cm}  -
\hspace*{ 1.15 cm}   T \ S_{\rm ep}(T,\bam_1,\bam_2)
\hspace*{ 0.5 cm}  -
\hspace*{ 0.5 cm} T \ {\cal{I}}(T,\bam_1,\bam_2)
\EEA

For the Hessian of $\beta F(T,m_1,m_2)/N$ 
we find the following expressions
\BEA
{\bf{H}}
&\equiv& \beta
\left(\begin{array}{c}
\hspace*{0.25cm} 
\frac{J^2 L^2}{w (w+T/2)^2}+T\frac{m_2+m_1^2}{(m_2-m_1^2)^2}
\hspace*{0.5cm} 
\frac{J^3 L}{2 w (w+T/2)^2}-T\frac{m_1}{(m_2-m_1^2)^2}\\
\\
\hspace*{0.25cm}\frac{J^3 L}{2 w (w+T/2)^2}-T\frac{m_1}{(m_2-m_1^2)^2}
\hspace*{0.5cm}
 \frac{J^4}{4 w(w+T/2)^2}+\frac{T}{2}\frac{1}{(m_2-m_1^2)^2}\hspace*{0.25cm}
\end{array}\right)
\\
&=&
\beta\frac{ J^2 }{2 w (w+T/2)^2}
\left(\begin{array}{c}
\hspace*{0.25cm} 
2 L^2
\hspace*{0.25cm} 
J L\\
\\
\hspace*{0.25cm}J L
\hspace*{0.25cm}
 \frac{J^2}{2}\hspace*{0.25cm}
\end{array}\right)+
\frac{1}{(m_2-m_1^2)^2}
\left(\begin{array}{c} 
m_2+m_1^2
\hspace*{0.5cm} 
-m_1\\
\\
\hspace*{0.25cm} -m_1
\hspace*{1cm}
 \frac{1}{2}
\end{array}\right)
\\
&=&  \hspace*{0.5 cm}{\mbox{Hessian of $\beta {\cal{H}}_{\rm eff}(m_1,m_2)$}}
 \hspace*{0.5 cm}
- \hspace*{0.5 cm} {\mbox{Hessian of ${\cal{I}}(m_1,m_2)$}} 
\EEA

The determinant of the Hessian of $\beta F(T,m_1,m_2)/N$, 
computed at equilibrium,  is
\BEQ
\det({\bf{H}})=\frac{1}{2(\bam_2-\bam_1^2)^3}
\left(1+Q_{\tinf}D+P_{\tinf}\right)
\EEQ
that is always positive. In the formula above  we introduce
the abbreviations
\BEA
&&Q\equiv \frac{J^2 (HJ+KL)}{\Kt^3 w\left( w+T/2\right)^2}
\hspace*{1cm},\hspace*{1cm}
Q_{\tinf}\equiv \lim_{t \to \infty} Q(m_1(t),m_2(t)) = Q(\bam_1,\bam_2),
\label{Q}\\
&&P\equiv \frac{J^4 (m_2-m_1^2)}{2 \Kt w \left( w+T/2\right)^2} 
\hspace*{1cm},\hspace*{1cm}
P_{\tinf}\equiv \lim_{t \to \infty} P(m_1(t),m_2(t)) = P(\bam_1,\bam_2).
\label{P}
\EEA
\noindent that we will often use in the following.
The inverse matrix turns out to be
\BEQ
{\bf{C}}\equiv{\bf{H}}^{-1}=
-\frac{\bam_2-\bam_1^2}{1+Q_{\tinf}D+P_{\tinf}}
\left(\begin{array}{cc}
\hspace*{-1.5cm} 
1+P_{\tinf}
\hspace*{2cm} 
2\bam_1-2\frac{L}{J}P_{\tinf}\\
\\
\hspace*{0.25cm}2\bam_1-2\frac{L}{J}P_{\tinf}
\hspace*{0.5cm}
4\bam_1^2+2(\bam_2-\bam_1^2)+4\frac{L^2}{J^2}P_{\tinf}\hspace*{0.25cm}
\end{array}\right).
\label{Hessian_inv}
\EEQ
The elements of this  matrix are the thermodynamic average 
of the fluctuations of $m_1$ and $m_2$ around their equilibrium values
$\bam_1$, $\bam_2$,
 times a factor $N$, as we can 
immediately chek expanding $F$ to the second order around
$\bam_1$ and $\bam_2$ in (\ref{Z_static}).
This holds for temperature high enough where, asymptotically,
the constraint (\ref{CONSTRAINT}) plays no role.

\renewcommand{\thesection}{\arabic{section}}
\section{Analytically solvable Monte Carlo dynamics with glassy aspects}
\setcounter{equation}{0}\setcounter{figure}{0}
\renewcommand{\thesection}{\arabic{section}.}
\label{sec3}

We assume as the dynamics a generalization of previously introduced
parallel Monte Carlo dynamics for the harmonic oscillators. 
This kind of analytic Monte Carlo approach was first introduced in
\cite{BPPR}, and later applied in ~\cite{BPR} to the simpler, exactly solvable
harmonic oscillator model (which is just our model after setting $J=L=0$)
and by one of us ~\cite{NPRL98,N00} also for a spherical spin model
(which is the present model after setting $H=K=0$ and considering the 
$\{x_i\}$ as quenched random variables).
 The thus obtained dynamical
model with a very simple Hamiltonian and a contrived dynamics has the
benefit of being not only programmable on a computer, but even being
solvable analytically, which yields a much deeper insight in its
properties. Moreover, in the long-time domain the dynamics looks
quite reasonable in regard to what one might expect of any system with a
VFTH-law in its statics.

In a Monte Carlo step a random updating of the variables is performed
($x_i\to x'_i=x_i+r_i/\sqrt{N}$) where the $\{r_i\}$ have a gaussian 
distribution with zero mean and variance $\Delta^2$.
We call $x$ (without any subscript) the energy difference 
between the new and the old state, viz.
$x\equiv{\cal{H}}(\{x'_i\})-{\cal{H}}(\{x_i\})$.
If the energy of the new configuration is higher than the 
energy of the initial configuration ($x>0$) the move is 
 accepted with a probability $W(\beta x)\equiv
\exp(-\beta x)$;
if the new energy is lower ($x<0$) it is accepted always ($W(\beta x)=1$).

The updating is parallel and it is this particular feature
giving the collective behaviour leading to exponentially divergent time scales
in a model with no  interactions between particles such us ours.
A sequential updating would not produce any glassy effect.
In this sense there is an analogy with facilitated Ising models \cite{ST96},
and with the kinetic lattice-glass model  with contrived dynamics
of Kob and Andersen \cite{KA2},
where the transition probabilities depend on the neighboring
configuration; this dynamics may induce glassy behavior in situations 
where ordinary Glauber dynamics \cite{GLAUBER}
would not. Models of these
types may give valuable insights in the long time dynamics, at least, 
within a class that exhibits some long-time universality.

In a Monte Carlo step the quantities $Nm_1=\sum_ix_i$ and 
$Nm_2=\sum_ix_i^2$ are updated. Let us denote their change by 
$y_1$ and $y_2$, respectively. Following \cite{N00}
we get the distribution function of $y_1$ and $y_{2}$, 
for given values   of $m_1$ and $m_2$,:
\BEA
p(y_1,y_2|m_1,m_2)&\equiv &\int\prod_i \frac{d r_i}{\sqrt{2\pi\Delta^2}}
\exp{\left(-\frac{r_i^2}{2\Delta^2}\right)}\,\,
\delta\left(\sum_i {x'}_i-\sum_ix_i-y_1\right)\,
\delta\left(\sum_i {x'}_i^{2}-\sum_ix_i^2-y_2\right)
\nn \\ &=&
\frac{1}{4\pi \Delta^2\sqrt{m_2-m_1^2}}
\exp\left(-\frac{y_1^2}{2\Delta^2}-\frac{(y_2-\Delta^2-2y_1m_1)^2}
{8\Delta^2(m_2-m_1^2)}\right)
\label{sec3:Ptrans}
\EEA
We can express the energy difference as 
\BEQ
x=\frac{\Kt}{2} \ y_2 -{\tilde{H}} \ y_1,
\label{sec3:x}
\EEQ
\noindent 
upon neglecting the variations of $m_1$ and $m_2$ that are
 of order $(y_k/N)^2\sim\Delta^2/N$.

In terms of the energy difference $x$
and of $y=y_1$ the distribution function can be formally written
 as the product of two other gaussian  distributions:
\BEA
p(y_1,y_2|m_1,m_2)dy_1 dy_2
	&=&\hspace*{ 15 mm}p(x|m_1,m_2)\hspace*{ 10 mm}p(y|x,m_1,m_2)
\hspace*{ 15 mm}dx \ dy\nn\\
&=&\frac{1}{\sqrt{2\pi \Delta_x}}
	\exp\left(-\frac{(x-{\overline{x}})^2}{2 \Delta_x}\right)
	\frac{1}{\sqrt{2\pi \Delta_y}}
	\exp\left(-\frac{(y-{\overline{y}}(x))^2}{2 \Delta_y}\right)
	dx \  dy
\label{sec3:PROBDIST}
\EEA

\noindent where 
\BEA
{\overline{x}}=\Delta^2{\tilde{K}}/2, \hspace*{2.5 cm} &&  \hspace*{2 cm}
\Delta_x=\Delta^2{\tilde{K}}^2
(m_2-m_1^2)+\Delta^2{\tilde{K}}^2\left(m_1-{\tilde{H}}/{\tilde{K}}\right)^2,\\
 {\overline{y}}(x)=\frac{m_1-{\tilde{H}}/{\tilde{K}}}{
m_2-m_1^2+\left(m_1-{\tilde{H}}/{\tilde{K}}\right)^2}
\frac{x-{\overline{x}}}{\tilde{K}},
 &&  \hspace*{2 cm}
\Delta_y=\frac{\Delta^2(m_2-m_1^2)}
{m_2-m_1^2+\left(m_1-{\tilde{H}}/{\tilde{K}}\right)^2}.
\label{def:MCave}
\EEA

The variance of the randomly chosen updating $\{r_i\}$
of the slow variables was in previous approaches 
\cite{BPR,NPRL98,N00,BPPR} a constant. That was enough to cause
 an Arrhenius relaxation of the glass. To find a VFTH-like relaxation,
in the present model  we let  $\Delta^2$ depend on the distance to the 
constraint, i.e. on  the whole $\{x_i\}$ configuration before the 
Monte Carlo step:

\BEQ
\Delta^2(t)\equiv 8[m_2(t)-m_1^2(t)]
\left(\frac{B}{m_2(t)-m_1^2(t)-m_0}\right)^\gamma
\label{sec3:Delta}
\EEQ
where $B$ is a constant and $\gamma$ is an exponent larger than zero
that we discussed already as being used in practice to make the best 
VFTH-type fitting of the relaxation time in experiments
\cite{ANGELL,MCKENNA}.
In our model $\gamma$ is a constant; it has no prescribed value
since we do not make any connection with a microscopic system.
In the standard VFTH-law one would just take $\gamma=1$. One of our
results will be to see that there are three qualitatively 
different regimes: $\gamma>1$, $\gamma=1$ and $0<\gamma<1$,
showing that the situation $\gamma=1$ is actually non-generic.

We also define a quantity that we shall frequently encounter in the
following,
\BEQ
\Gamma(t)\equiv \left(\frac{B}{m_2(t)-m_1(t)^2-m_0}\right)^\gamma.
\label{sec3:Gamma}
\EEQ
The nearer the system goes to the constraint (i.e. the smaller the 
value of $m_2-m_1^2-m_0$), the larger the variance becomes, 
thus implying almost always a refusal of the proposed updating.
In this way, in the neighborhood of the constraint,  the dynamics 
is very slow  and goes on through  very seldom but very
large moves, a thing that can be interpreted as activated dynamics.
When the constraint is reached the $\Gamma$ becomes infinite and the system
 dynamics is stuck forever. 
The system does not evolve anymore towards equilibrium
 but it is blocked in one  single ergodic component
of the configuration space. At large enough temperatures,
the combination $m_2(t)-m_1^2(t)-m_0$ will remain strictly positive.
The highest temperature, $T_0$, at which it can vanish for
$t\to\infty$, is identified with the Kauzmann temperature $T_K$.

The question whether detailed balance  is satisfied or not 
is also non-trivial  in our model. Indeed, 
it happens to be satisfied for this kind of dynamics
 only for large $N$. For exact detailed balance we should have 
\BEQ
p(x|m_1,m_2)\exp(-\beta x)=p(-x|m_1,m_2)
\EEQ
\noindent
but now, when we perform the inverse move $\{x'_i\}\to\{x_i\}$,
the probability distribution is also depending on the $\{r_i\}$
through $\Delta^2$ as defined in (\ref{sec3:Delta}). Thus the right hand
 side of the detailed balance  consists of
a $p(-x|m'_1,m'_2;\Delta'^2)\neq p(-x|m_1,m_2;\Delta^2)$.
Expanding this probability distribution in powers of $1/N$, however,
we get that $ p(-x|m'_1,m'_2;\Delta'^2)= p(-x|m_1,m_2;\Delta^2) 
+\cO(\Delta^2/N)$.
Other terms of $\cO(\Delta^2/N)$ were already neglected in the approximation
of $x$ done in (\ref{sec3:x}).
  So, inasmuch as the whole approach is
valid only for $N\to\infty$, detailed balance is also satisfied;
it would slightly be violated in a finite $N$ simulation. 
We work at very large $N$
and, even though $\Delta^2\propto\Gamma(t)$ grows as 
the system approaches equilibrium 
(it even diverges at the Kauzmann  temperature),
 we perform first the  thermodynamic
limit computing the dynamics equation and only  eventually
the  limit
$t\to \infty$. If we would do the opposite there would be a region around 
the Kauzmann temperature where the detailed balance is violated and the 
dynamics is not the one discussed here. 
However, this is not our aim since we 
are interested in the ergodicity breaking that takes place in systems 
with a large number  (Avogadro like)  of variables.

In the harmonic oscillator model and in the spherical spins model studied in 
\cite{BPR,NPRL98,N00} the dynamics was performed
within this approach, but at fixed $\Delta$.
Both cases showed a relaxation time diverging 
at low temperature with an Arrhenius law, typical of {\em strong} glasses.
We could also study enhanced Arrhenius law by setting $m_0=0$ in the present
 model but here we want, instead, to develop a model representing a 
{\em{fragile}} glass with a Kauzmann transition at a finite temperature.

The Monte Carlo  equations for the dynamics of $m_1$ and $m_2$ can now
be derived according to the lines of ~\cite{N00}. They read: 
\BEA
&&\dot{m}_1=\int dy_1dy_2 W(\beta x)\ y_1 \ p(y_1,y_2|m_1,m_2) =
\int dx W(\beta x)\  {\overline{y}}(x) \ p(x|m_1,m_2)\ \ ,
\label{eq:m1}\\
&&\dot{m}_2=\int dy_1dy_2  W(\beta x)\ y_2 \ p(y_1,y_2|m_1,m_2) =
\frac{2}{\tilde{K}}\int dx W(\beta x) \ 
 (x+{\tilde{H}}\  {\overline{y}}(x)) \ p(x|m_1,m_2) \ .
\label{eq:m2}
\EEA

Before performing the study of dynamics
 we define two new variables $\mu_1$ and $\mu_2$
depending on $m_1$ and 
$m_2$ and representing, respectively, the deviation from the equilibrium 
state and the distance from the constraint:
\BEA
&&\mu_1\equiv\frac{\tilde{H}}{\tilde{K}}-m_1 \ \ ,\\
&&\mu_2\equiv m_2-m_1^2-m_0.
\EEA
When $\mu_1=0$ equilibrium is obtained and the equilibrium value of $m_1$
is given by the solution of the equation
\BEQ
\frac{\tilde{H}(\bam_1,\bam_2)}{\tilde{K}(\bam_1,\bam_2)}=\bam_1 \ .
\label{sec3:eqm1}
\EEQ
\noindent that is the saddle point equation (\ref{SPeq1}), with $\bam_k$
being the equilibrium values of $m_k$, and 
 it can be proven equivalent to a fourth order  equation.

When $\mu_2=0$ the constraint is reached. This will happen if the temperature
is low enough ($T\leq T_0$). $T_0$ is the highest temperature at which
the constraint is finally reached by the system.

Above $T_0$ equilibrium will be achieved without reaching the constraint. 
The temperature is too high for the system to notice that there is a 
constraint at all on the  configurations: 
$\lim_{t\to \infty}\mu_2(t)= \bamu(T)>0$.

At and below $T_0$ the system goes to configurations that become arbitrarily
close to the constraint, and then stays there arbitrarily long.

When all system parameters are fixed (aging setup)
the equations of motion 
(\ref{eq:m1})-(\ref{eq:m2}) become in terms of $\mu_1$ and $\mu_2$ 

\BEA
\dot{\mu_1}&=&-J Q \int dx  \ W(\beta x)\  x \  p(x|m_1,m_2) 
-(1+ Q D) \int dx \ W(\beta x)\  {\overline{y}}(x)\   p(x|m_1,m_2)
\label{sec3:mu1dot}
\\
\dot{\mu_2}&=&\frac{2}{\tilde{K}}\int dx\  W(\beta x)\  x   p(x|m_1,m_2) 
+2 \mu_1  \int dx \ W(\beta x)\  {\overline{y}}(x) \  p(x|m_1,m_2)
\label{sec3:mu2dot}
\EEA
\noindent where we have used $D$ and $Q$ defined respectively
in (\ref{D}) and (\ref{Q}).

We also shorten the expression ${\tilde{K}}(m_2-m_1^2)$ by the parameter
\BEQ
T_e\equiv\Kt (m_2-m_1^2) \ ,
\EEQ
\noindent  possibly depending on time through $m_1(t) $ and $m_2(t)$.
For the moment this is just an  abbreviation
  but in the next section we will show
that an alternative description of the dynamics is possible where
$T_e(t)$ turns out to be a mapping of the history of the system into 
an effective thermodynamic parameter. 
This 
 {\em {effective temperature}} would be  the temperature  of a system 
at equilibrium
visiting with the same frequency the same states 
that the actual - out of equilibrium - system at temperature $T$
is visiting 
 on a given time-scale during its dynamics.

In the  time regime
where $\Gamma \gg x^2/T_e^2 \sim {\cal{O}}(1)$ ($\mu_2(t) \ll 1$),
 the gaussian distribution
of the $x$ can be approximated by 
\BEQ
p(x|m_1,m_2)\simeq \frac{\exp \left(-\Gamma\right)}
{4T_e\sqrt{\Gamma \pi}} \ \ 
\exp\left(\frac{x}{2 T_e}\right)\ \ \left(1-\frac{x^2}{16T_e^2\Gamma}
+\frac{x^4}{512 T_e^4\Gamma^2}\right)
\EEQ

\noindent and the equations (\ref{sec3:mu1dot}),  (\ref{sec3:mu2dot}) become
\BEA
\dot{\mu_1}&=&4 \Upsilon \left[J Q {\tilde{K}} (m_0+\mu_2) r  
\left(1-\frac{3(1-2r+2r^2)}{\Gamma}\right)
-\mu_1(1+QD)\left(\Gamma-(1-3r+4r^2)\right)
\right] \label{appeq:MU1}\\
\dot{\mu_2}&=&-4 \Upsilon\left[2(m_0+\mu_2) r
\left(1-\frac{3(1-2r+2r^2)}{\Gamma}\right)
- \mu_1^2 \left(\Gamma-(1-3r+4r^2)\right)\right]\label{appeq:MU2}
\EEA
\noindent where $r$ is the normalized difference between  the 
parameter $T_e$ and the heat-bath
temperature $T$:
\BEQ
r\equiv\frac{T_e-T}{2T_e-T} 
\label{def:r}
\EEQ
and
\BEQ
\Upsilon\equiv  \frac{\exp(-\Gamma)}{\sqrt{\pi\Gamma}}(1-r).
\label{sec3:upsilon}
\EEQ
$\Upsilon$ (upsilon)
is the leading term of the expansion of the integral representing the 
acceptance rate of the Monte Carlo dynamics
\BEQ
\int dx W(\beta x) p(x|m_1,m_2)
 \simeq \frac{\exp(-\Gamma)}{\sqrt{\pi\Gamma}}
(1-r)\left[1-\frac{1}{2\Gamma}
(1-2r+4r^2)+ {\cal{O}}(\mu_2^{2\gamma})\right]. \label{def:INTEGRALE}
\EEQ

The solutions to the equations (\ref{appeq:MU1}) and (\ref{appeq:MU2})
depend on the relative size  of $\mu_1$ and
$\mu_2$, thus also on $\gamma$,
 as well as on $r$ that has a different behaviour above 
$T_0$, where $T_e$ tends to $T$ in the infinite time limit,
and below, where $T_e$  never equals the heat-bath temperature 
(see section \ref{sec4}).

The solution to the equation (\ref{appeq:MU2})
can be easily found neglecting
the second term, proportional to $\mu_1^2$.
It is expressed in the implicit form
\BEQ
2 \pi\frac{ \mbox{erf}(i\Gamma (t))}{i} - 2
\frac{\exp (\Gamma(t))}{\Gamma(t)^{1/\gamma}}=
 \frac{8 r_1 m_0 \gamma}{\pi}  t +\mbox{const}
\label{sol:MU2exact}
\EEQ
\noindent
where 
\BEQ 
\mbox{erf}(z)=\frac{2}{\sqrt{\pi}}\int_0^{z}e^{-t^2} dt.
\EEQ

To second order approximation this can be written as:
\BEQ
\mu_2(t)\simeq \frac{1}{\left[\log (t/t_0)+
c\log\left(\log(t/t_0)\right)\right]^{1/\gamma}}
\label{sol:MU2}
\EEQ 
The constants $t_0$ and $c$  depend on the temperature phase as will
be clarified in the following.

Above $T_0$ (\ref{sol:MU2}) is more precisely  the  behaviour of 
$\delta\mu_2(t)\equiv\mu_2(t)-\bamu(T)$. Since in this range of temperature
$T_e(t)-T\sim\delta\mu_2(t)$,
the first order expansion of $r$ is:
\BEQ
r\simeq \frac{\delta\mu_2(t)}{m_0+\bamu(T)}
\left(1+\frac{P_{\tinf}}
{1+Q_{\tinf}D}\right)-
\mu_1(t)\frac{2P_{\tinf}D}{J\Kt_{eq}(m_0+\bamu(T))(1+Q_{\tinf}D)}
\label{exp:r}
\EEQ
\noindent where $P_{\tinf}$ and $Q_{\tinf}$ are given in 
(\ref{P})-(\ref{Q}).
In this case in equation (\ref{sol:MU2}) $c$ is equal to $1/2$ and 
 the expression of $t_0$ in terms of the
parameters of the model is:

\BEQ
t_0\equiv \frac{\sqrt{\pi}}{8\gamma \Gamma(0)[1+P_{\tinf}/(1+Q_{\tinf}D)]}
\EEQ
\noindent where $\Gamma(0)$ is the initial value of $\Gamma(t)$.

\begin{figure}[!htb]
\epsfxsize=11cm
\epsfysize=7cm
\centerline{\epsffile{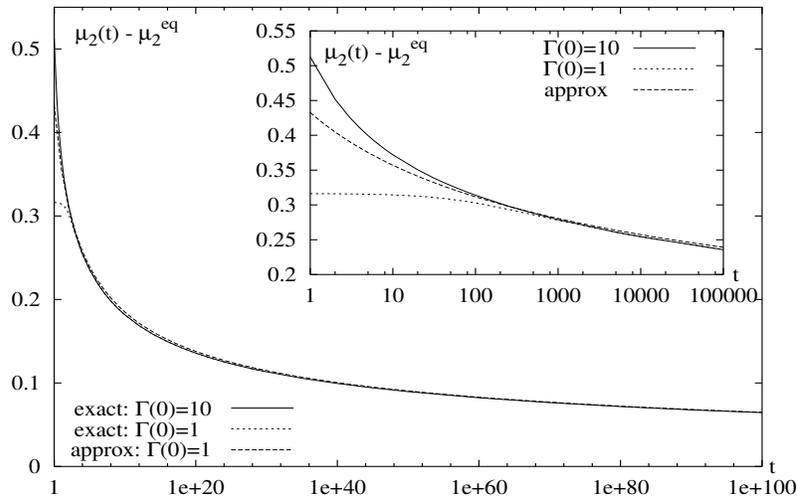}}
 \protect\caption{\small{The difference between $\mu_2(t)$ and its asymptotic
value $\bamu(T)$ is plotted for heat bath
temperature $T=0.41$, slightly above
 the Kauzmann temperature  $T_0=4.00248$.
The case is plotted with $K=J=1$, $H=L=0.1$, $m_0=5$. 
The upper two curves represent the exact solution (\ref{sol:MU2exact})
with two 
different initial conditions.
The lower one is the  approximated solution (\ref{sol:MU2}).
In the inset the initial behaviour  is shown:
clearly the approximation 
is valid already after a few decades of the dynamics.}}
\label{fig:mu2_t}
\end{figure}

Below $T_0$ the qualitative behaviour of 
$\mu_2(t)$ (in this case the $\bamu$ part
is zero) is the same, but $T$ is never reached. 
This implies
that $r$ goes to some asymptotic constant $r_{\tinf}$.
Concerning  the solution (\ref{sol:MU2}) the only difference is in the values
\BEQ
c=\frac{2+\gamma}{2\gamma} \ \ ;\hspace*{2 cm}
t_0\equiv \frac{B \sqrt{\pi}}{8m_0\gamma r_{\tinf}(1-r_{\tinf})}.
\EEQ

In figure \ref{fig:mu2_t} we show the exact solution, 
numerically computed, of 
equation (\ref{appeq:MU2}) for a particular choice of the parameter values:
$K=J=1$, $H=L=0.1$, $m_0=5$, $B=1$, $\gamma=2$. We can see that after a couple
of decades the behaviour is the one given in (\ref{sol:MU2}).

The ratio of equations (\ref{appeq:MU1}) and (\ref{appeq:MU2}) brings the 
equation:
\BEQ
\frac{d\mu_1}{d\mu_2}=\frac{\mu_1(1+QD)(\Gamma+2-3r+2r^2)-JQT_er}
{2r(m_0+\mu_2)-\mu_1^2(\Gamma+2-3r+2r^2)}
\label{eq:dmu1dmu2}
\EEQ

With respect to the relative weight of $\mu_1$ and 
$\mu_2$ we can identify different regimes, where the solution has
different behaviours.

\begin{enumerate}
\item{$T>T_0$. \ }
The leading term of the solution is given by  the stationary solution.
We can also neglect the term of $\cO(\mu_1^2\Gamma)$ in the denominator.
Using the expansion (\ref{exp:r}) for $r$ we get::
\BEQ
r\simeq \frac{1}{m_0+\bamu}\frac{1+P_{\tinf}+Q_{\tinf} D}{1+Q_{\tinf} D}
\ \delta\mu_2
\EEQ
 and
\BEQ
\mu_1(t)= 
\frac{T J Q_{\tinf}\left(1+P_{\tinf}+Q_{\tinf}D\right) }
{(m_0+\bamu)(1+Q_{\tinf}D)^2}\frac{\delta\mu_2(t)}{\baG} 
+{\cal{O}}(\delta\mu_2^{2})+\cO(\delta\mu_2^{2\gamma+1})\label{sol:mu11}
\EEQ
\noindent Here we have also expanded $\Gamma(t)$
as
\BEQ
\Gamma(t)=\baG-\gamma\baG
\frac{\delta\mu_2(t)}{\bamu}.
\EEQ
\noindent We are most interested in what happens
next to the  the Kauzmann temperature, i.e. for  very big $\baG$,
 at long but not extremely long times, that means $\delta\mu_2(t)$
 small but not vanishing.
A more detailed treatment, including an expansion in $T-T_0$ of
 $\bamu$ appearing in $P_{\tinf}$ and $Q_{\tinf}$, 
can also be done,
looking carefully up to which extent $\bamu$ can be approximately
neglected with respect to  $\delta\mu_2(t)$.
We can neglect $\bamu$ with respect to the whole $\mu_2(t)$
at temperatures very close to the Kauzmann temperature and
 for times that are not extremely long, so that  we are far from 
the thermalization and the dynamics has still aging behaviour.
In figure (\ref{fig:ratio}) we show the relative weight of 
$\bamu$ on the whole $\mu_2(t)$ for a specific case.
As  is clear from the figure, as soon as we go too far from $T_0$, 
we cannot neglect with respect to $\mu_2(t)$ its asymptotic value
$\bamu$.

\begin{figure}[!htb]
\epsfxsize=11cm
\epsfysize=7cm
\centerline{\epsffile{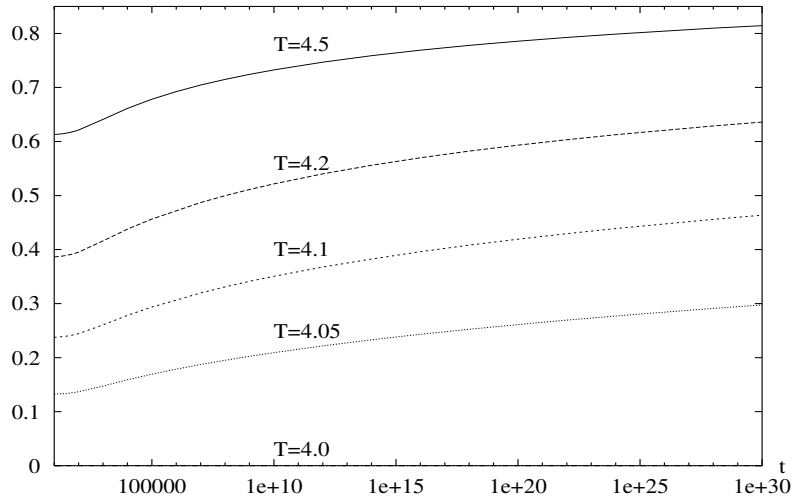}}
 \protect\caption{\small{Ratio of $\bamu(T)/\mu_2(t)$ at different 
temperatures, at and above the Kauzmann temperature. Too far away
from $T_0$ the contribution of $\bamu(T)$ to $\mu_2(t)$ becomes
relevant. The case is plotted with $K=J=1$, $H=L=0.1$, $m_0=5$. 
For this set of parameters the Kauzmann temperature turns out to be
$T_0=4.00248$ }}
\label{fig:ratio}
\end{figure}

\item{$T<T_0$, $\gamma>1$. \ }
In this and in the following cases the asymptotic value
of $\mu_2(t)$ is $\bamu=0$ so that
$\delta\mu_2(t)=\mu_2(t)$.
 Also in this dynamic regime 
the adiabatic approximation can be carried out and the second term
in the denominator of (\ref{eq:dmu1dmu2}) is again negligible.
In this case the leading term of $r$ in its expansion in powers of $\mu_2$,
$r_{\tinf}$,
is of $\cO(1)$. Therefore we get 
\BEQ
\mu_1(t)=\frac{J\baTe r_{\tinf}Q_{\tinf}}
{1+Q_{\tinf}D}\frac{1}{\Gamma(t)}
+{\cal{O}}(\mu_2^{1+\gamma}) \ ,
\EEQ
\noindent where 
\BEA
&&\baTe \equiv\lim_{t\to \infty} 
{\tilde{K}}(m_1(t),m_2(t))(m_0+\mu_2(t))=\Kt_{\tinf} m_0 \ ,
\\
&&
\Kt_{\tinf}\equiv \Kt\left(\bam_1,\bam_2\right),
\label{Ktinf}
\\
&&
r_{\tinf}=\frac{\baTe -T}{2\baTe -T}.
\label{rinf}
\EEA

\item{$T<T_0$, $\gamma=1$.}
In this case the adiabatic expansion is no more consistent. We have to
solve the equation (\ref{eq:dmu1dmu2})
taking $d\mu_1/d\mu_2$ into account. To leading order
the equation takes the form:
\BEQ
\frac{d\mu_1}{d\mu_2}=\frac{
\mu_1\Gamma(1+Q_{\tinf} D)-JQ_{\tinf}\baTe r_{\tinf}}
{2r_{\tinf}m_0}+\cO(\mu_1)+\cO(\mu_2)+\cO(\Gamma\mu_1^2)+\cO(\Gamma\mu_1\mu_2).
\EEQ
Defining the quantity $\epsilon\equiv\frac{B(1+Q_{\tinf}D)}
{ 2r_{\tinf}m_0}$
we identify other five sub-regimes in the case $\gamma=1$.
\begin{enumerate}
\item{ $\epsilon>1$.}
 The solution is 
\BEQ
\mu_1(t)=\frac{JQ_{\tinf}\baTe r_{\tinf}}{2r_{\tinf}m_0(\epsilon-1)}
\mu_2(t)
-c_1\frac{1}{\epsilon-1}
\mu_2^{\epsilon}(t) .
\label{sec3:mu1_31}
\EEQ
The exponent $\epsilon$
is always positive, at least in cooling, because 
$\baTe >T$ making  $r_{\tinf}$ and $Q_{\tinf}$ positive.
$c_1$ is also positive because it is the exponential  of the integration 
constant (the value of which depends on the initial conditions).
Since $\epsilon>1$, the second term in the right hand side can be neglected
and $\mu_1\sim\mu_2$.

\item{ $\epsilon=1$.} 
We find:
\BEQ
\mu_1(t)=-\frac{J\baTe r_{\tinf}Q_{\tinf}}{1+Q_{\tinf}D}\frac{\log\mu_2(t)}
{\Gamma(t)}+c_2 \mu_2,
\EEQ
where $c_2$ is the integration constant and can take any value.
In the long time dynamics  the logarithm term will take over
and, independently from the initial conditions,
 will be $\mu_1 > \mu_2$ and positive.

\item{ $1/2 <\epsilon<1$.}
The second term in (\ref{sec3:mu1_31}) is leading and the solution is
\BEQ
\mu_1(t)=c_1\frac{1}{1-\epsilon}
\mu_2^{\epsilon}(t).
\label{sec3:mu1_32}
\EEQ
$c_1$ is a positive constant and $\mu_1\gg\mu_2$ and  positive.

\item{$\epsilon=1/2$.}
When
If $\epsilon\leq 1/2$ the second term in the denominator, always
neglected up to now, has to be taken into account.
In this case the leading term in the denominator goes to zero and 
$J\baTe r_{\tinf}Q_{\tinf}$ can be
neglected with respect to $\mu_1\Gamma(1+Q_{\tinf}D)$ in the numerator.
We can thus easily solve the equation:
\BEQ
\frac{d\mu_2}{d\mu_1}=\frac{2r_{\tinf}m_0-2\mu_1^2\Gamma}
{\mu_1\Gamma(1+Q_{\tinf} D)}.
\label{sec3:eqmu1inv}
\EEQ
For $\epsilon=1/2$ we get
\BEQ
\mu_2(t)=-\frac{2}{1+Q_{\tinf}D}\mu_1^2(t)\log\mu_1(t)+c_2\mu_1^2(t)
\label{sec3:mu1_34}
\EEQ
 that is not invertible analytically. It is clear anyway that
in this sub-regime $\mu_1\gg\mu_2$. $c_2$ can take any value.

\item{$\epsilon<1/2$.}
The solution is
\BEQ
\mu_1(t)=\sqrt{\frac{m_0r_{\tinf}(1-2\epsilon)}{\Gamma(t)}}
\left(1+\frac{c_1}{2}\left(\frac{1+Q_{\tinf}D}{2}\right)^{1/\epsilon}
\left(\frac{1-2\epsilon}{\epsilon}\right)^{1/\epsilon-1}
\mu_2(t)^{1/2\epsilon-1}\right)
\label{sec3:mu1_35}
\EEQ
where $c_1>0$. It is still $\mu_1\gg\mu_2$.
\end{enumerate}

\item{$T<T_0$, $\gamma<1$.}
Considering  also the term  $\mu_1^2\Gamma$ in the denominator of
(\ref{eq:dmu1dmu2}), the solution
 is now:
\BEQ
\mu_1(t)=\sqrt\frac{r_{\tinf}m_0}{\Gamma(t)}\left(1-
\frac{1+Q_{\tinf}D}{2m_0r_{\tinf}\gamma}\mu_2(t)\Gamma(t)\right).
\label{sec3:mu1_4}\EEQ
In this low temperature regime is $\mu_1\gg\mu_2$ once again.

\end{enumerate}

For $\gamma=1$, $\epsilon\leq 1/2$ and for $\gamma<1$ the solution
to equation (\ref{sec3:eqmu1inv}) involves only
 the absolute value of $\mu_1$, thus giving two possible choices for the 
sign of the function $\mu_1(\mu_2)$. 
In order to  guarantee continuity of $\mu_1$ at the parameters values at
which the dynamics changes regime, we imposed to the $\mu_1$
in two contiguous regimes  to have the same sign. That means that in 
(\ref{sec3:mu1_34}), (\ref{sec3:mu1_35}) and (\ref{sec3:mu1_4})
we chose the plus sign. 

The time dependent variables $\mu_1(t)$, $\mu_2(t)$ give the dynamic
behaviour of every observable in the long, but not extremely long,
 time regime, i. e. in the aging regime.
When the time increases further the dynamic will exponentially relax 
to equilibrium like $\exp(-t/\tau_{eq})$. We will see what $\tau_{eq}$
is in the next section.

\renewcommand{\thesection}{\arabic{section}}
\section{Out of Equilibrium Thermodynamics}
\setcounter{equation}{0}\setcounter{figure}{0}
\renewcommand{\thesection}{\arabic{section}.}
\label{sec4}

The history of a system that is  far
 from equilibrium can be expressed by a number of
effective parameters, like the 
{\it effective temperature}
 or other {\it effective fields}, in order to
recast the out of equilibrium dynamics in a thermodynamic approach\cite{N00}.
 The number of effective parameters needed to make such a translation
is, in principle, equal to the number quantities considered.
For a certain class of systems, however, there is some effective
 thermalization and the effective parameters  pertaining to 
processes having the same time scale become asymptotically 
equal to each other in time.
Examples of out of equilibrium regimes governed by 
a single effective temperature have been considered in \cite{N00}\cite{N2}.
In computer glasses the approach applies with some success.
\cite{SKT,CR1}. 

Given the solution of the dynamics (thus the form of the
 functions $m_1(t)$ and $m_2(t)$) a quasi-static approach 
can be followed by  computing the 
 partition function $Z_e$ of all the macroscopically 
equivalent states (those having the same values for $m_{1,2}$) 
at the given time $t$.
The measure on which this out of equilibrium partition function
is carried on is not the Gibbs measure.
In order to generalize the equilibrium thermodynamics we 
assume an effective temperature $T_e$ and an effective field $H_e$,
and substitute the equilibrium measure by 
$\exp (-\H_{\rm eff}(\{x_i\},T,H_e)/T_e)$, 
where $\H_{\rm eff}$ is introduced in (\ref{def:Heff})
 and the true external field $H$ in it has been substituted by
an effective field $H_e$.
$T_e$ and $H_e$ are at this step of the computation just fictitious 
parameters. 
However, as soon as  we get the expression of the ``thermodynamic''
potential $F_e\equiv-T_e \log Z_e$  as a function
of macroscopic variables $m_{1,2}$ and effective parameters,
we can fix $T_e$ and $H_e$ as taking those values that make 
the potential as small as possible.
 We thus have to minimize $F_e$ with
respect to $m_1$ and $m_2$ to determine $T_e$ and $H_e$ and evaluate
the resulting analytic expressions at $m_1=m_1(t)$ and 
$m_2=m_2(t)$ given by the dynamics at the considered time $t$.
Counting all the 
macroscopically equivalent states at the time $t$, at which the
dynamical variables take values $m_1$ and $m_2$,
we get:
\BEQ
Z_e\left(m_1,m_2;T_e,H_e\right)\equiv \int {\cal{D}}x 
\ \exp\left[-\frac{1}{T_e}\H_{\rm eff}(\{x_i\},T,H_e)
\right] \ 
\delta( N m_1-\sum_ix_i)\ \delta( N m_2-\sum_ix_i^2),\label{Ze}
\EEQ

From this partition function we can build an effective thermodynamic potential
as a function of $T_e$ and $H_e$, besides of $T$ and $H$, where
the effective parameters depend on time through the
time dependent values of $m_1$ and $m_2$ solutions of the dynamics. 
They actually are
a way of describing the evolution in time of the system out of equilibrium.
The effective free energy takes the form:
\BEQ
F_e(t)=U\left(m_1(t),m_2(t)\right)
-T\S\left(m_1(t),m_2(t)\right)-T_e(t)\I\left(m_1(t),m_2(t)\right)
+[H-H_e(t)]Nm_1(t), \label{Fe}
\EEQ
\noindent with 
\BEA
T_e(t)&=&\Kt(t) (m_0+\mu_2(t)),
 \label{sec4:Te}
\\
 H_e(t)& =&H-\Kt \mu_1(t) \ .
\label{sec4:He}
\EEA
\noindent where the last term of $F_e$ replaces the $-HNm_1$ occurring
in $U$ (see eq. (\ref{def:U})) by $-H_eNm_1$, and where
\BEQ
\I(t)=\frac{N}{2}\left\{1+\log\left[m_0+\mu_2(t)\right]\right\}
\label{sec4:I}
\EEQ
\noindent is the configurational entropy and where $U$ and $\S$ are given
in (\ref{def:U}) and (\ref{def:Sep}).

As we see from (\ref{sec4:Te}) and (\ref{sec4:He})
in the dynamic regimes $1$ and $2$, reported in section
\ref{sec3}, where $\mu_1 \ll \mu_2$, the effective temperature alone is
enough for a complete thermodynamic description of the dominant
physic phenomena ($H_e=H$), while in the regimes $3 a, 3b$ ($\mu_1\sim \mu_2$)
and in $3 c, 3d, 3e$ and $4$ ($\mu_1\gg \mu_2$), when $\mu_1$ becomes no more 
negligible, 
 the effective field $H_e$ is also needed.

\subsubsection{Effective temperature  from generalized first law}
Calling $M\equiv Nm_1$, and using (\ref{def:U}), (\ref{def:Sep}) 
and (\ref{def:I}),
the differential of the free energy (\ref{Fe}) turns out to be:
\BEQ
dF=-S_{\rm ep} dT-\I dT_e-M dH_e\ ,
\EEQ
\noindent thus implying
\BEQ
dU=T dS_{\rm ep} + T_e d\I+(H_e-H) dM -MdH\ .
\EEQ
Using this expression we are able to write down the
first law of thermodynamics $dU=\dbarrm Q + \dbarrm W$, in the 
two temperature-two fields case,
 where
the change in work done on the system is,  $\dbarrm W=-MdH$.
In order for the conservation of energy to be satisfied
the heat variation has, then,  to take the form 
\BEQ
 \dbarrm Q = T dS_{\rm ep} + T_e d\I + (H_e-H) dM \ .
\label{sec4:dQ}
\EEQ
This is the same expression obtained in the two temperature picture of 
\cite{N2} where the fields where absent.
At equilibrium, where $H_e=H$ and $T_e=T$,
this reduces to the usual expression for ideal reversible quasi-static
transformations $ \dbarrm Q = T dS$, with the total
entropy $S=S_{\rm ep}+\I$.

From (\ref{sec4:dQ}) the complete expression for the rate of change of 
the heat of the system turns out to be:
\BEQ
\dot Q=\frac{T\Kt^2(w+T/2)}{2DJ}N\dot\mu_1+\frac{\Kt}{2}N\dot\mu_2+
\frac{\Kt\mu_1}{2}\frac{N}{1+QD-\Kt J Q\mu_1}
\left(\dot\mu_1+\dot\mu_2\Kt JQ\right)
\EEQ
The heat flowing out of the system is $-\dot Q$.
Referring to the aging regimes 
described in section \ref{sec3}
the quantity $\dot Q$ 
 turns out to be proportional to $\dot\mu_2$ in the regimes $1$ ($T>T_0$)
and $2$ ($T<T_0, \gamma>1$).
In the dynamic regimes $3a$ and $3b$ ($T<T_0, \gamma=1, \epsilon\geq1$) 
is $\dot Q\propto \dot\mu_1+\dot\mu_2$.
For $3c$, $3d$, $3e$ ($T<T_0, \gamma=1$, $\epsilon\leq 1$) and for 
regime $4$ ($T<T_0, \gamma<1$) $\dot Q\propto \dot\mu_1$.

In every dynamic regime
 $\dot\mu_1$ and 
 $\dot\mu_2$ are negative and this implies that the heat flow of the
out of equilibrium system 
is  positive in its approach to equilibrium, as it should,
no matter the values of the 
parameters of the model.

Starting from the first law of thermodynamics,
we can derive  the effective temperature in yet another way, 
through a generalization of  the Maxwell relation $T=\p U/\p S$ 
valid at equilibrium
for a system of internal energy $U$ and entropy $S$, with the derivative taken 
at constant magnetization (or volume).
We put for simplicity $H_e=H$ in the rest of this subsection.
Out of equilibrium,
together with the previous Maxwell relation for 
equilibrium processes (where $S$ has to be substituted by $S_{\rm ep}$)
 also the following generalization
holds:
\BEQ
T_e=\left.\frac{\p U}{\p \I}\right|_{S_{\rm ep}} \ \ .
\EEQ
A more feasible identity, where the variable to be kept constant
during the transformation is the bath temperature, rather than the
entropy of the fast processes, can be obtained 
\cite{FV,CR2,MP}. 
 Let's introduce with this aim the function $\Phi$:
\BEQ
\Phi\equiv F_e+T_e \I
\EEQ
inducing $d\Phi=T_ed\I-S_{\rm ep}dT$.
Through this auxiliary potential function 
$\Phi$ we can then rewrite the effective temperature as
\BEQ
T_e(t)=\left.\frac{\partial \Phi}{\p \I}\right|_T
\label{sec4:TeMax}
\EEQ
This result is a firm prediction for systems that satisfy the assumption of
a two temperature thermodynamics. For underlying mechanisms in specific cases
see \cite{FV,CR2,MP}.
Writing the latter  as $\dot{\Phi}/\dot{\I}$ and using (\ref{eq:m1}) and 
(\ref{eq:m2}) we get (neglecting terms of $\cO(\mu_1)$):
\BEQ
T_e(t)=\Kt(m_1(t),m_2(t))\left[m_2(t)-m_1^2(t)\right]
\EEQ
\noindent
 in agreement with  (\ref{sec4:Te}).

\subsection{Statics}

$T_0$ is defined as the temperature at which the constraint is reached 
from above: some configurations become infeasible and the valleys
of the free energy landscape are divided by infinite barriers. 
The breaking of the ergodicity in a landscape with many 
minima gives rise
to a real thermodynamic phase transition \cite{KAUZ}.

When the constraint (\ref{CONSTRAINT}) on the phase space of the 
$\{x_i\}$ is first reached, at $T_0$, in the infinite time limit, 
$\I$ goes to its minimal value $\I_0\equiv \I(T_0)=1+\log{m_0}$.
Zero configurational entropy would mean that only one configuration 
is allowed  for the system.
Coming from high temperature there would  thus be a transition from a many
(meta-stable) states phase to a phase in which the system is
stuck forever in one single minimum.
This transition is what is thought to happen in real glasses, at the so called
Kauzmann temperature.
Since we are using the continuous variable $\{x_i\}$, 
the entropy  ${\cal{I}}$ (as well as $S_{\rm ep}$) 
is, in our case,  ill defined at low temperature: 
 it would diverge like $\log T$ at zero temperature if
no constraint would be present. 
Our value ${\cal{I}}(T_0)$
is greater than zero,
because this entropy counts all the multiple ways in which the continuous
harmonic  oscillators can arrange themselves in order to satisfy the 
constraint 
(\ref{CONSTRAINT}). 
Since we are dealing with classical variables we
 can bypass this inconvenience just subtracting from $\I$
the constant $\I_0$ to make $\I(T=T_0)=0$. The entropy value
$\I_0$ is related to dynamics on time scales where all the degenerate
minima are sampled. These are much longer than the scales of our
interest, and for our purposes the constant $\I_0$ plays no role.

\begin{figure}[!htb]
\epsfxsize=10cm
\centerline{\epsffile{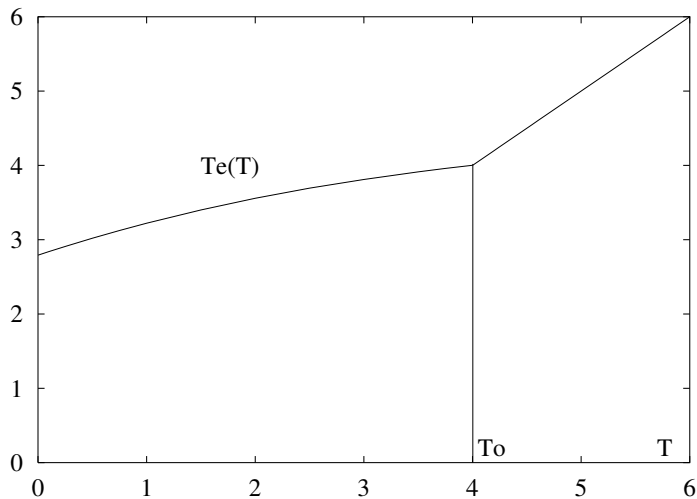}}
 \protect\caption{\small{In the static regime the 
effective temperature is shown as a function of the heat bath temperature.
At high temperature they coincide but below the Kauzmann transition
$T_e$ does not reach $T$, not even in the infinite time limit.
The system remains out of equilibrium for ever. 
Values of the constants are $K=J=1$, $H=L=0.1$, $m_0=5$}}
\label{fig:Te_T}
\end{figure}

To see how the  transition takes place we first look at the asymptotic 
behaviour of the effective temperature.
When $T\geq T_0$ and $t\to \infty$, $T_e$ becomes 
the heat bath temperature $T$.
When $T<T_0$, instead, $T_e$ never reaches such a temperature.
It rather goes towards some limit value $\baTe(T)$ that we can get 
from the equation (\ref{sec4:Te}). It may be
 rewritten for clearness in the explicit form
\BEQ
(Km_0-\baTe+T)(Km_0-\baTe)(J\baTe)^2
+D^2m_0(Km_0-\baTe)^2
-J^4m_0 (\baTe)^2=0
\label{eq:TeT}
\EEQ
\noindent a quartic equation for the effective temperature 
in the infinite time limit.
The same equation evaluated at $T_e=T=T_0$ gives us the value of 
the Kauzmann temperature $T_0$ 
as a function of the parameter of the model.
In figure (\ref{fig:Te_T})  we plot $T_e$ versus $T$ for a choice 
of parameter values.

From (\ref{eq:TeT}) or Figure \ref{fig:Te_T}
we observe that $\left.dT_e/dT\right|_{T_0^{-}}<1$
whereas, coming from above the Kauzmann temperature,
one has, of course, $\left.dT_e/dT\right|_{T_0^{+}}=1$.
The derivative of $T_e(T)$ shows thus a discontinuity at $T=T_0$.
Any thermodynamic function, like $U$ and $m_1$, will 
depend on the heat bath temperature both explicitly and through this
effective temperature.
For the specific heat we will have, for instance:
\BEQ
C\equiv\left.\frac{1}{N}\frac{d U}{d T}\right|_H=
\left.\frac{1}{N}\frac{\p U}{\p T}\right|_H + 
\left.\frac{1}{N}\frac{\p U}{\p T_e}\right|_{H,T}
\left.\frac{d T_e}{d T} \right|_H.
\EEQ
This is of the same form $C=c_1+c_2\left(\p T_e/\p T\right)_p$
 assumed originally by Tool\cite{TOOL} for the study of 
caloric behaviour in the glass formation region.

The  discontinuity in $\left.d T_e/d T \right|_{H}$
causes a discontinuity in the specific heat and also in the 
quantity $-\left.\p m_1/\p T\right|_H$, called magnetizability in 
\cite{NPRL98,N00} (it is 
 the analogue of a thermal expansivity for 
the model here described),  
 because both of these quantities
contains terms proportional to $\left.\p T_e/\p T\right|_H$.
One could now discuss the Ehrenfest relations between
these discontinuities, and the Prigogine-Defay ratio, as was done for
related models by one of us~\cite{N2,NPRL98,N00}. Because of 
the close analogy between all these cases, we shall not go deeper into
this at the present moment.

\subsection{Dynamics}

The relaxation time is the characteristic time on which
the system initially out of equilibrium (because, for instance, of 
a sudden quench to low temperature) relaxes towards equilibrium.
It can be defined, for instance, from the dynamical equations of the 
internal energy per harmonic oscillator $u\equiv U/N$
\BEQ
\dot{u}=-\frac{u}{\tau_{eq}} \ ,
\label{def:taueq}
\EEQ
\noindent or, 
equivalently, from the equations
of motion for $m_1$, $m_2$ as the time at which  the quantity of interest goes
to $1/e$ of its initial value.
In any temperature regime comes out that the relaxation time
has an exponential behaviour in $\Gamma$:

\BEQ
\tau_{eq}\sim e^{\Gamma} = \exp\left(\frac{B}{\mu_2}\right)^{\gamma}
\label{sec4:tau}
\EEQ
Making use of the solution (\ref{sol:MU2}) we find the following behaviour 
for the relaxation time versus the heat bath temperature:

\begin{enumerate}
\item{$T>T_0$. \ }
$\mu_2(t)\to\bamu(T)$ and near enough to the Kauzmann temperature we 
can linearize the latter
in $T-T_0$. For $t\to\infty$ we get an exponential decay with
relaxation time
\BEA
&&\tau_{eq}\propto\exp\left(
\frac{A_0}{T-T_0}\right)^{\gamma}
\label{sec4:VFTH}
\\
&&A_0=B\,\left(\left.\frac{\p \mu_2^{eq}(T)}{\p T}\right|_{T_0}\right)^{-1}=
\left.\frac{ B \Kt_{\tinf}(K-\Kt_{\tinf})(1+DQ_{\tinf}+P_{\tinf})}
{(K-\Kt_{\tinf})(1+DQ_{\tinf})-\Kt_{\tinf}P_{\tinf}}\right|_{T_0}
\EEA
This behaviour is a generalized Vogel-Fulcher-Tammann-Hesse (VFTH) law
\cite{VFTH1,VFTH2,VFTH3}, where $\gamma$ can have any value and in particular 
$\gamma=1$.
Looking at the configurational entropy,
since at the first order expansion in $\mu_2$ we have 
${\cal{I}}-{\cal{I}}_0\simeq\frac{N}{2m_0}\mu_2$, we also find from 
(\ref{sec4:tau}) the  Adam-Gibbs relation \cite{AG}:
\BEQ
\tau_{eq}\propto \exp\left[\frac{N B}
{2 m_0\left({\cal{I}}-{\cal{I}}_0\right)}\right]^{\gamma}
\label{VFTHeq}
\EEQ
Far from equilibrium, in the aging regime where the relaxation
is very slow, we can still define a time dependent ``relaxation time''
giving the characteristic time scale on which the $\alpha$ processes are 
taking place.
Always for $T$ very near to $T_0$, in the aging regime, $\bamu$, the static part of $\mu_2$,
is negligible with respect to the dynamic part $\delta\mu_2$ so that
for the effective temperature we have the following expansion:
\BEQ
T_e(t)\simeq T_0+\frac{1+DQ_{\tinf}+P_{\tinf}}{1+Q_{\tinf}D}\Kt_{\tinf}\delta\mu_2(t)
+\cO(T-T_0).
\EEQ
 we get
\BEA
&&\tau(t)\propto \exp\left(\frac{A(T)}
{T_e(t)-T}\right)^\gamma\simeq\exp\left(\frac{A(T_0)}
{T_e(t)-T_0}\right)^\gamma
\label{VFTHag}
\\
&&A(T)\equiv\frac{B(1+DQ_{\tinf}+P_{\tinf})}{1+Q_{\tinf}D}\Kt_{\tinf}
\label{A}
\EEA
where $A(T_0)< A_0$, meaning that in the static $\tau$ is more divergent.

\item{$T<T_0, \gamma>1$. \ }For $T<T_0$ the relaxation time always
diverges
for $t\to\infty$. 
However,  as it was done  in the case above $T_0$ for the relaxation
in the aging regime,  an  instantaneous 
relaxation time  can be considered and
expressed in terms of the effective temperature
using the first order expansion of $T_e$ in $\mu_2$:
\BEQ
T_e(t)=\baTe+\frac{1+DQ_{\tinf}+P_{\tinf}}{1+Q_{\tinf}D}\Kt_{\tinf}\mu_2(t).
\EEQ
We find 
\BEQ
\tau(t)=\tau(T,T_e(t))\propto\exp\left(\frac{A(T)}
{T_e(t)-\baTe(T)}\right)^\gamma
\label{VFTHlow}
\EEQ
\noindent where $A(T)$ is the one written in (\ref{A}).
 The aging behaviour just above
 and well below $T_0$ are thus intimately related.
The expression (\ref{VFTHlow}) resembles a VFTH law  where  the heat bath temperature
has been substituted by a time dependent effective temperature $T_e(t)$
and the Kauzmann temperature by the asymptotic value $\baTe$.
Such a relation for the time scale of the aging dynamics
could hold very well in more general systems.

\item{$T<T_0, \gamma\leq1$. \ } In these regimes, where $\mu_1$ cannot be
 neglected with respect to  $\mu_2$, there is no simple expression for $\tau$.

\end{enumerate}

\renewcommand{\thesection}{\arabic{section}}
\section{Two-time  variables: breaking of time-translation invariance 
and the fluctuation-dissipation relation}
\setcounter{equation}{0}\setcounter{figure}{0}
\renewcommand{\thesection}{\arabic{section}.}
\label{sec5}

In this section we compute the  correlation
and response functions that, unlike the energy and the quantities
 $m_1(t)$ and $m_2(t)$,
depend in a 
non-trivial way on two times, when the system is out of equilibrium,
 thus showing directly the loss of Time Translation Invariance (TTI)
with respect 
to the case at equilibrium.
The aim of computing such quantities 
is also to build a Fluctuation Dissipation relation and look
at the meaning of the Fluctuation-Dissipation Ratio (FDR), 
$\p_{t'}C(t,t')/G(t,t')$,  far from equilibrium.

The correlation functions between the thermodynamic
fluctuation of a quantity $m_a(t)$ at time $t$
and that of a quantity $m_b(t')$ at a different time $t'$
are defined like:
\BEQ
C_{ab}(t,t')\equiv N \left<\delta m_a(t)\delta m_b(t')\right>, 
\ \ \ \ a, b=1, 2
\EEQ
\noindent where $\left< .... \right>$ is the average over 
the dynamic processes, i. e. the harmonic oscillators.

The response of an observable $m_a$ at time $t$ to a perturbation in 
a conjugate field $H_b$ at some previous time $t'$ takes the form:
\BEQ
G_{ab}(t,t')\equiv\frac{\delta\left<m_a(t)\right>}{\delta H_b(t')},
\ \ \ \ a, b=1, 2
\EEQ
\noindent In our model $H_1=H$ and $H_2$=$K/2$.

Since we will very often  make use of the derivatives with respect
 to $m_1$ and $m_2$
of the integrals given by
 the Monte Carlo dynamics introduced in section \ref{sec3},
 we show them in appendix explicitly computed
and we shorten the notation  defining
the variables $f_k$ and $g_k$, $k=0,1,2,3$ in (\ref{def:f0})-(\ref{def:g3}).


\subsection{High temperature case: $T>T_0$}

First we analyze the case above the Kauzmann temperature.
In this case the expansion of the 
(\ref{def:f0})-(\ref{def:g3})
in powers of  $\mu_2(t)$ becomes both an expansion in $\delta\mu_2(t)$ and
  in  $\bamu$ (or equivalently in $1/{\overline{\Gamma}}
= \left(\bamu /B\right)^{\gamma}$), 
 because we are interested in studying 
what happens for large times and 
 near the Kauzmann temperature $T_0$, i.e. for small values of
$\delta\mu_2(t)$and for small values 
of $\bamu$ (or large values of $\baG$).
In the $f_k$ and $g_k$ written in appendix
 we left the notation without 
the bar meaning that an approximation for long  times,
 i.e. an expansion in $\delta\mu_2(t)$, or an expansion in $\bamu$,
has still to be done, 
depending on the kind of approximation that we need. 
For the sake of clearness we repeat here the expansion of $\Gamma(t)$
in the aging regime:
\BEQ
\mu_2(t)=\bamu +\delta\mu_2(t)\hspace*{1cm};\hspace*{2 cm}
\Gamma(t)=\baG-\gamma\baG\frac{\delta\mu_2(t)}{\bamu }.
\EEQ

The following exact relations hold:
\BEQ
\hspace*{-1 cm}\partial_{m_1}\mu_1=-1-LQ{\tilde{K}}, \hspace*{2cm}
\partial_{m_2}\mu_1=-\frac{J}{2}Q{\tilde{K}}\\
\EEQ

We stress that $\p_{m_1} \mu_1$ and $\p_{m_2} \mu_1$
are still functions of $\mu_2$, through $Q$ and $\Kt$, and that they can
thus 
be expanded in  powers of $\delta\mu_2$, leading to corrections 
to the $f_k$ and $g_k$.
Since in the end they will only appear in the combinations
$\p_{m_1}\mu_1+2m_1\p_{m_2}\mu_1$ and $\p_{m_1}\mu_1-2L/J\p_{m_2}\mu_1$
 we just give the expressions of these ones:
\BEA
&&\partial_{m_1}\mu_1+2m_1\partial_{m_2}\mu_1=-(1+QD)\simeq-(1+Q_{\tinf}D)
-\delta\mu_2 Q_1 D,
\\
&&\partial_{m_1}\mu_1-2 \frac{L}{J}\partial_{m_2}\mu_1=-1,
\EEA
\noindent where
\BEQ
Q_1=\frac{Q_{\tinf}}{(1+Q_{\tinf}D)(m_0+\bamu )}
\left[\frac{J^2(m_0+\bamu )(3 {\overline{w}}+T/2)}
{2 {\overline{w}}^2( {\overline{w}}+T/2)}-3P_{\tinf}\right]
\label{Q_1}
\EEQ
is the coefficient of $\delta\mu_2(t)$
in the  first order expansion of $Q$:
\BEA
Q(t)=Q_{\tinf}+Q_1\delta\mu_2(t)
\label{Qexp} .
\EEA
In the following formulae the derivatives
of $\mu_1$, as well as $\mu_1$ itself,
 have to be considered as general, regular functions of $\mu_2$.

Besides the expansion (\ref{Qexp}) of $Q$
we also give the expansions to first order in $\delta\mu_2(t)$ of 
the quantities $r$ (defined in (\ref{def:r})), $m_1$, $\Kt$ (\ref{def:tilde}), 
$P$ (\ref{P}) and $Q$ (\ref{Q}):
 \BEA
&&r(t)=\frac{1}{m_0+\bamu }\left(\frac{1+Q_{\tinf}D+P_{\tinf}}
{1+Q_{\tinf}D}\right)\delta\mu_2(t)
\label{rexp}
\\
&&
\Kt(t)=\Kt_{\tinf}+\frac{\Kt_{\tinf}P_{\tinf}}{(1+Q_{\tinf}D)(m_0+\bamu )}
\delta\mu_2(t)
\\
&&m_1(t)=\bam_1-\frac{P_{\tinf} D}{\Kt_{\tinf}J(1+Q_{\tinf}D)(m_0+\bamu )}
\delta\mu_2(t)
\\
&&P(t)=P_{\tinf}+\frac{P_{\tinf}}{(1+Q_{\tinf}D)(m_0+\bamu )}
\left[1+Q_{\tinf}D-P_{\tinf}+\frac{J^2(m_0+\bamu )(3 {\overline{w}}+T/2)}
{2 {\overline{w}}^2( {\overline{w}}+T/2)}\right]\delta\mu_2(t)
\label{Pexp}
\EEA
\noindent where ${\overline{w}}\equiv\sqrt{J^2\bam_2+2JL\bam_1+L^2+T^2/4}$.
For what concerns the terms containing $\Gamma\mu_1$, 
from (\ref{sol:mu11}) we saw
that in this dynamic regime 
\BEQ
\Gamma\mu_1=\frac{J Q_{\tinf} T}{m_0+\bamu}
\frac{1+Q_{\tinf}D+P_{\tinf}}{(1+Q_{\tinf}D)^2}\delta\mu_2+
\cO\left(\delta\mu_2^2\right)
\EEQ
\noindent

In  \cite{N00} equations of motion for simpler models were obtained.
Our present model share the basic attributes that are needed 
to get those equations, namely
the possibility of writing 
the
transition probability (\ref{sec3:Ptrans})
of the Monte Carlo  dynamics as the product of 
two gaussian probability distributions (\ref{sec3:PROBDIST}), 
functions respectively of the energy variation $x$
and of the variation $y$ of the magnetization like quantity $\sum_i x_i$. 

We thus recall here 
the following equations holding  
for the equal time correlation functions:
\BEA
\frac{d}{dt} && C_{ab}(t,t)=
\int W(\beta x)\left\{{\overline{y}}_a(x){\overline{y}}_b(x)
+\Delta_y\left(-\frac{H_1}{H_2}\right)^{a+b-2}\hspace*{-0.9 cm}
+\sum_{c=1,2}\frac{\p \ }{\p m_c}
\left[{\overline{y}}_a(x) C_{cb}(t,t)+{\overline{y}}_b(x) C_{ca}(t,t)\right]
\right\} p(x|m_1,m_2) \ dx, 
\\
\nn
&&\hspace*{15 cm} a,b=1,2.
\EEA
\noindent where (recalling (\ref{sec3:x}), (\ref{def:MCave})
and using (\ref{sec3:Delta}) and (\ref{sec4:Te}))
\BEA
&&{\overline{y}}_1(x)=\frac{\mu_1}{m_2-m_1^2+\mu_1^2}\frac{{\overline{x}}-x}
{\Kt}= \left(4\Gamma\mu_1-\mu_1\frac{x}{T_e}\right)+\cO(\mu_1^3)
\\
&&{\overline{y}}_2(x)=\frac{2}{\Kt}\left(x+\Ht\ {\overline{y}}_1(x)\right)
\\
&&{\overline{x}}=\frac{\Delta^2\Kt}{2}=4T_e\Gamma 
\\
&&\Delta_y=\frac{\Delta^2(m_2-m_2)}{m_2-m_1^2+\mu_1^2}=
8(m_0+\mu_2)\Gamma+\cO(\Gamma\mu_1^2)
\EEA
Expanding the integrals  $\int W(\beta x)
{\overline{y}}_a(x){\overline{y}}_b(x)p(x|m_1,m_2)  dx$ and   
$\Delta_y\int W(\beta x)p(x|m_1,m_2) dx$
 up to  order $\Upsilon$, the equations become:
\BEA
\dot{C}_{11}(t,t)&=& 8 \left[m_0+\mu_2(t)\right] \Gamma(t) 
\left(1-\frac{1-2r+4r^2}{\Gamma(t)}\right)\Upsilon(t)+
2 f_1(t) C_{11}(t,t)+2 g_1(t) C_{12}(t,t)\ ,
\label{eq:C11tt}
\\
\dot{C}_{12}(t,t)&=& 16 m_1(t) \left[m_0+\mu_2(t)\right] \Gamma(t) 
\left(1-\frac{1-2r+4r^2}{\Gamma(t)}\right)\Upsilon(t)+
\label{eq:C12tt}
\\
\nn
&&\hspace*{8 cm}+
f_2(t) C_{11}(t,t)+ \left[f_1(t)+g_2(t)\right] C_{12}(t,t)+
g_1(t) C_{22}(t,t)\ ,
\\
\dot{C}_{22}(t,t)&=&32 m_1(t)^2 \left[m_0+\mu_2(t)\right] \Gamma(t) 
\left(1-\frac{1-2r+4r^2}{\Gamma(t)}\right)\Upsilon(t)+ 
32\left[m_0+\mu_2(t)\right]^2\Upsilon(t)+
\label{eq:C22tt}
\\
\nn
&&\hspace*{11 cm}+
2 f_2(t) C_{12}(t,t)+2 g_2(t) C_{22}(t,t) \ .
\EEA

Due to the complicated form of the equations we are not able to find 
solutions valid at every time. We are obliged to find approximate solutions
valid on given time scales.
First we will study the solutions in the aging regime, for times that
are long but
 not longer than some
given time-scale, $t_g$, after which the system begins to thermalize.
Afterwards we will study the dynamics of correlation and response functions 
for times much longer than $t_g$, while
the system is approaching to equilibrium.

\subsubsection{Dynamics in the aging regime}
In the aging regime, for temperature just above the Kauzmann temperature $T_0$,
we can neglect $\bamu$ with respect to $\delta\mu_2(t)$.
This means that in expressions (\ref{Qexp})-(\ref{Pexp}) we have
to put $\bamu$ equal to zero  everywhere, including the 
constants $Q_{\tinf}$, $P_{\tinf}$ and $\Kt_{\tinf}$ (defined respectively in 
(\ref{Q}), (\ref{P}) and (\ref{Ktinf}))
 and we  can write $\delta\mu_2(t)=\mu_2(t)$.

To find the solutions to (\ref{eq:C11tt})-(\ref{eq:C22tt}) we can firstly 
 perform an adiabatic expansion neglecting
the time derivatives of the correlations functions.
Indeed, to first order of approximation $\dot{C}_{ab}$ is  
proportional to $\dot\mu_2 $: they are of 
${\cal{O}}(\delta\mu_2\Upsilon)$, negligible
with respect to the right hand side terms.
We then compute the second order corrections.
The  solutions for the case $T>T_0$, in the aging regime  with negligible
$\bamu$ and $r$ proportional to  $\mu_2(t)$, turn out to be:

\BEA
C_{11}(t,t)&=&
\frac{1}{1+Q_{\tinf}D}
\left\{m_0+\mu_2(t)\left[1-\frac{m_0Q_1D}{1+Q_{\tinf}D}\right]
+\cO\left(\frac{1}{\Gamma}\right)+\cO(\mu_2^2(t))\right\}
\label{sec5:C11tt}
\\
C_{12}(t,t)&=&
\frac{1}{1+Q_{\tinf}D}
\left\{2m_1(t)m_0+\mu_2(t)\left[
2m_1(t)\left(1-\frac{m_0Q_1D}{1+Q_{\tinf}D}\right)
-m_0Q_{\tinf}D \right]
+\cO\left(\frac{1}{\Gamma}\right)+\cO(\mu_2^2(t))
\right\}
\label{sec5:C12tt}
\\
C_{22}(t,t)&=&\frac{1}{1+Q_{\tinf}D}
\left\{4m_1(t)^2 m_0+\mu_2(t)\left[
4m_1(t)^2\left(1-\frac{m_0Q_1D}{1+Q_{\tinf}D}\right)
-4 m_0 m_1Q_{\tinf}D \right]
+\cO\left(\frac{1}{\Gamma}\right)+\cO(\mu_2^2(t))
\right\}
\label{sec5:C22tt}
\EEA
To get these expressions it is enough to keep in $f_1$, $g_1$, $f_2$, $g_2$ 
(defined in (\ref{def:f1})-(\ref{def:g2})) only terms up to $\cO(\Upsilon)$.

Once we have the equal time solutions we can solve the equations for the two
times functions.
Always following the approach of \cite{N00} we get the equations:
\BEQ
\p_t C_{ab}(t,t')=f_a(t) C_{1b}(t,t')+
g_a(t) C_{2b}(t,t')\hspace*{2 cm} a, b= 1, 2.
\label{sec5:Cab}
\EEQ
\noindent where $f_a$ and $g_a$ are defined in (\ref{def:f1})-(\ref{def:g2}).
We introduce  the function
\BEA
{\tilde{f}}&\equiv& f_1+\frac{2\dot{m_1} g_1}{g_2-2 m_1 g_1}-g_1\frac{f_2-2m_1f_1}{g_2-2 m_1 g_1}
\label{ft}
\\ 
\nn
&=& -4\Upsilon\left\{(1+QD)\Gamma-\left[1+QD-\frac{2 DQP}{1+QD}-\frac{DP(1+QD)}{\gamma(1+P+QD)}\right]
+\cO\left(\frac{1}{\Gamma}\right)\right\}
\EEA
where $\dot{m}_1$ is obtained from (\ref{eq:m1}) as
\BEQ
\dot{m}_1\simeq4\mu_1\Upsilon\Gamma-4\mu_1\Upsilon(1-3r+4r^2)=\cO(\mu_2\Upsilon)
\EEQ
and is negligible with respect to the leading orders.

The decoupled equations for $C_{11}$ and $C_{12}$ that we get are, 
in this notation:
\BEQ
\p_t C_{1b}(t,t')= {\tilde{f}}(t) \ C_{1b}(t,t'),
\hspace*{4 cm} b=1,2 \ .\label{eq:C1b}
\EEQ
\noindent To the leading order
 the two correlation functions above 
are connected to $C_{21}$ and $C_{22}$ in the following
way:
\BEQ
C_{2b}(t,t')\simeq 2 m_1(t) C_{1b}(t,t'), \hspace*{4 cm} b=1,2 \ .
\EEQ

Defining the time evolution function for the considered  time-scale sector as
\BEQ
{\tilde{h}}(\tau)\equiv\exp\left(-\int_0^{\tau}{\tilde{f}}(t)dt\right)
\EEQ
\noindent  the solution of (\ref{eq:C1b}) comes out to be
\BEQ
C_{1b}(t,t')=C_{1b}(t',t')\frac{\tilde{h}(t')}{\tilde{h}(t)}+\cO(\mu_1\Upsilon)
\EEQ

Following the approach of 
\cite{N00} we also  derive the response function.
Neglecting the terms of  ${\cal{O}}(\Upsilon^2)$ (called {\em
switch terms} in \cite{N00}) they are:
\BEA
G_{11}(t,t^{\tp})&=&-\beta\int dy_1dy_2 W'(\beta x) y_1^2 p(y_1,y_2|m_1,m_2)=
-\beta \int dx W'(\beta x) \left[{\overline{y}}_1(x)^2 +\Delta_y\right]
p(x|m_1,m_2)=
\\ \nonumber
&=&
-\beta\int dx W'(\beta x)\Delta_y p(x|m_1,m_2)+\cO(\mu_2^2 \Upsilon)
\simeq 
\frac{4\Upsilon\Gamma}{\Kt}-\frac{2\Upsilon}{\Kt}+\cO(\mu_2 \Upsilon)
\label{sec5:G11_tt}
\\
G_{12}(t,t^{\tp})&=&-\beta\int dy_1dy_2 W'(\beta x) y_1 y_2 p(y_1,y_2|m_1,m_2)=
\\ \nn &=&
-\beta \frac{2}{\tilde{K}}\int dx W'(\beta x) 
\left\{x{\overline{y}}_1(x)+{\tilde{H}}
\left[{\overline{y}}_1(x)^2 +\Delta_y\right]\right\}
p(x|m_1,m_2)=
\\ \nonumber
&=&
-2\beta\frac{\tilde{H}}{\tilde{K}}
\int dx W'(\beta x)\Delta_y p(x|m_1,m_2)+
\cO\left(\frac{\mu_2}{\Gamma}\Upsilon\right)
\hspace*{0.4 cm}\simeq
\frac{8m_1}{\Kt}\Upsilon\Gamma-\frac{4m_1\Upsilon}{\Kt}+
\cO\left(\mu_2\Upsilon\right)
\label{sec5:G12_tt}\\
G_{22}(t,t^{\tp})&=&-\beta\int dy_1dy_2 W'(\beta x) y_2^2 p(y_1,y_2|m_1,m_2)=
\\ \nn &=&
-\beta \frac{4}{\tilde{K}^2}\int dx W'(\beta x) 
\left\{x^2+2{\tilde{H}} x {\overline{y}}_1(x)+{\tilde{H}}^2
\left[{\overline{y}}_1(x)^2 +\Delta_y\right]\right\}
p(x|m_1,m_2)=
\\ \nonumber
&=&
-4\beta\frac{1}{\tilde{K}^2}
\int dx W'(\beta x)\left(x^2+\tilde{H}^2\Delta_y\right) p(x|m_1,m_2)
+\cO\left(\frac{\mu_2}{\baG}\Upsilon\right)
\hspace*{0.3 cm}\simeq
\frac{16 m_1^2}{\Kt}\Upsilon\Gamma-\frac{8m_1^2\Upsilon}{\Kt}-32\Upsilon m_0^2
+\cO\left(\mu_2\Upsilon\right).
\label{sec5:G22_tt}
\EEA

The equations describing the evolution in $t$ of the response to
a perturbation at $t'$ have the same shape of 
 those  for the correlation functions 
(\ref{sec5:Cab}).
 The solutions are then
\BEQ
G_{ab}(t,t')=G_{ab}(t',t')\frac{\tilde{h}(t')}{\tilde{h}(t)}.
\label{sec5:Gab}
\EEQ

With these results we can generalize the Fluctuation Dissipation Theorem
(FDT) defining another effective temperature, $T_e^{FD}$,
by means of the ratio between the derivative with respect to the 
initial time (also called  ``waiting'' time) $t'$ of the
correlation function $C_{11}$ and the response function $G_{11}$:

\BEQ
T_e^{FD}(t,t')\equiv\frac{\partial_{t'}C_{11}(t,t')}{G_{11}(t,t')}.
\EEQ

To compute it we need:
\BEA
\partial_{t'}C_{11}(t,t')&=&\left(\p_{t'}C_{11}(t',t')\right)
\frac{\tilde{h}(t')}{\tilde{h}(t)}
-{\tilde{f}}(t')
\frac{\tilde{h}(t')}{\tilde{h}(t)}C_{11}(t',t')\simeq
-{\tilde{f}}(t')
\frac{\tilde{h}(t')}{\tilde{h}(t)}C_{11}(t',t')
\\
\nn
&\simeq&
-4\Upsilon(t')\left\{(1+Q_{\tinf}D+Q_1D\mu_2(t'))\Gamma(t')+\cO(1)\right\}
\\
\nn
&\times&
\frac{1}{1+Q_{\tinf}D}
\left[m_0+\mu_2(t)\left(1-\frac{m_0Q_1D}{1+Q_{\tinf}D}\right)
\right]\frac{\tilde{h}(t')}{\tilde{h}(t)}
\\
\nn
&\simeq&
-4\Upsilon(t')\left\{\left[m_0+\mu_2(t')\right]\Gamma(t')+\cO(1)\right\}
\EEA

Eventually we get
\BEQ
T_e^{FD}(t,t')\simeq T_e(t')
\left[1+\cO\left(\frac{1}{\Gamma(t')}\right)
+\cO\left(\mu_2(t')^2\right)
\right]=T_e^{FD}(t')
\label{sec5:TeFD}
\EEQ
where $T_e $ was first introduced in section \ref{sec3} and later on 
derived in (\ref{sec4:Te}).
We recall  that $\Gamma^{-1} \propto \mu_2^{\gamma}$.
As we see here the above defined fluctuation dissipation effective temperature
coincides,  in the time-scale of our interest, with the effective temperature $T_e$
that we got by the quasi static approach,
 only if $1/\Gamma$
is negligible with respect to
$\mu_2$.
This is true only if $\gamma$, the exponent of the generalized 
VFTH law (\ref{sec4:VFTH}), is greater than one. 
Otherwise the last correction
is no more sub-leading: already for $\gamma=1$, $T_e^{FD}\to T_e$
only in the infinite time limit, i. e. for time-scales  longer than those
of the considered aging regime.
As already discussed in section \ref{sec3}, where we presented 
 the results of the dynamics of the one-time  observables,
the value of the exponent  $\gamma$ discriminates between different regimes . 
For $\gamma>1$  an out of equilibrium thermodynamics can be built in terms of 
a single additional effective parameter (the effective temperature $T_e$).
For $\gamma \leq 1$, $T_e$ alone does not give consistent 
 results
in the generalization of the equilibrium properties to the non equilibrium 
case and in order to cure this inconsistency more effective parameters
are probably needed.
This discrepancy was clear, from section \ref{sec3}, for the regimes below 
$T_0$ where already the one-time variables had different behaviours
depending on the value of $\gamma$ being  greater, equal to  or lesser than $1$.
For $T>T_0$ there was not such a difference at at the one-time level.
It shows up, instead, at the level of 
two-time observables, as we just saw.

\subsubsection{Approach to equilibrium}
For times longer than the aging regime time-scales the terms
 that are relevant in the equations
(\ref{eq:C11tt})-(\ref{eq:C22tt}) for the correlation functions and 
in the expressions  (\ref{sec5:G11_tt})-(\ref{sec5:G22_tt}) for 
the response functions
are different. When $t\gg t_g$ the equilibrium value
$\bamu$ of the variable $\mu_2$ is no more negligible with respect to its 
time dependent part $\delta\mu_2(t)$
(that is eventually going to zero as $t \to \infty$).
We are in a regime  where $r\simeq 0$ ($T_e\simeq T$).
In the resolution of the equations (\ref{eq:C11tt})-(\ref{eq:C22tt})
this means that all the terms of $\cO(r \Upsilon \gamma/\mu_2)=
\cO(\delta\mu_2\Upsilon)$ 
are now sub-dominant  with
respect to those of $\cO(\Upsilon \Gamma)$ and of $\cO(\Upsilon)$.
To solve equations (\ref{eq:C11tt})-(\ref{eq:C22tt}) we use the  adiabatic 
expansion, as in the previous case.

The solutions for very large time, with finite, though small $\bamu$ and a vanishing $r$,
are: 
\BEA
C_{11}(t,t)&=&\frac{m_0+\mu_2(t)}
{1+QD+P}(1+P)
+ c_{11,r} \ r
 \label{sec5:C11}
\\
\nn
&=&\frac{m_0+\bamu}
{1+Q_{\tinf}D+P_{\tinf}}(1+P_{\tinf})+
c_{11,\delta\mu_2}\ \delta\mu_2(t)
\\
\hspace*{-0.5 cm}C_{12}(t,t)&=&C_{21}(t,t)\simeq
\frac{m_0+\mu_2(t)}
{1+QD+P}\left(2\m_1(t)-2\frac{L}{J}P\right)
+c_{12,r} \ r
=
\label{sec5:C12}
\\
\nn
&=&\frac{m_0+\bamu}
{1+Q_{\tinf}D+P_{\tinf}}\left(2\bam_1-2\frac{L}{J}P_{\tinf}\right)+
c_{12,\delta\mu_2}\ \delta\mu_2(t)
\\
C_{22}(t,t)&\simeq&
\frac{m_0+\mu_2(t)}{1+QD+P}\left[4m_1(t)^2
+2\left(m_0+\mu_2(t)\right)+4\frac{L^2}{J^2}P\right]
+ c_{22,r}\ r
\\
\nn
&=&\frac{m_0+\bamu}{1+Q_{\tinf}D+P_{\tinf}}\left[4\bam_1^2
+2\left(m_0+\bamu\right)+4\frac{L^2}{J^2}P_{\tinf}\right]
+ c_{22,\delta\mu_2}\ \delta\mu_2(t)
\label{sec5:C22}.
\EEA

Where
\BEA
c_{11,r}&\equiv&-\frac{4\gamma (m_0+\mu_2(t))^3 \Kt^2 J^2 Q^2   }
{\mu_2(t)(1+DQ)(1+P+DQ)^2}(1+P)
\\
c_{12,r}&\equiv&\frac{4 \gamma (m_0+\mu_2(t))^3\ \Kt J  Q}
{\mu_2(1+DQ)(1+P+DQ)^2}
\left[(1+P)(1+\Kt L Q)+ \Kt J Q\left(P \frac{L}{J}-m_1\right)\right]
\\
c_{22,r}&\equiv&\frac{16 \gamma (m_0+\mu_2(t))^3\  \Kt J Q}
{\mu_2(t)(1+DQ)(1+P+DQ)^2}
\left(P \frac{L}{J}-m_1\right)\left(1+\Kt L Q\right)
\EEA
and
\BEA
c_{11,\delta\mu_2}&\equiv&\frac{1}{1+DQ_{\tinf}+P_{\tinf}}
\left\{1+P_{\tinf}-(m_0+\bamu)\frac{P_1 DQ_{\tinf}+Q_1 D(1+P_{\tinf})}
{1+P_{\tinf}+DQ_{\tinf}}\right.
\\
\nn
&&\hspace*{10 cm}
-\left.\frac{8\gamma DQ_{\tinf}P_{\tinf}(1+P_{\tinf})(m_0+\bamu)}
{\bamu(1+DQ_{\tinf})^2}\right\}
\\
c_{12,\delta\mu_2}&\equiv&\frac{2}{1+DQ_{\tinf}+P_{\tinf}}
\left\{\left[\bam_1-\frac{L}{J}P_{\tinf}
\left(1+\frac{D}{L\Kt_{\tinf}(1+DQ_{\tinf})}\right)\right]\right.
\\
\nn
&&\hspace*{6 cm}-\left.
\frac{(m_0+\bamu)\left[
P_1\left(\bam_1+\frac{L}{J}
(1+DQ_{\tinf})\right)+DQ_1\left(\bam_1-\frac{L}{J}\right)\right]}
{1+P_{\tinf}+DQ_{\tinf}}\right.
\\
\nn
&&\hspace*{6 cm}+\left.
\frac{4\gamma DP_{\tinf}
\left[
(1+P_{\tinf})(1+\Kt_{\tinf}LQ_{\tinf})
+\Kt_{\tinf}JQ_{\tinf}(L/J P_{\tinf}-\bam_1)\right]}
{J\Kt_{\tinf}\bamu(1+DQ_{\tinf})^2}\right\}
\\
c_{22,\delta\mu_2}&\equiv&\frac{2}{1+DQ_{\tinf}+P_{\tinf}}
\left\{2(m_0+\bamu)+2 \bam_1^2+2\frac{L^2}{J^2}P_{\tinf}-
\frac{4\bam_1P_{\tinf}D}{\Kt_{\tinf}J(1+DQ_{\tinf})}\right.
\\
\nn
&-&\left.
\frac{(m_0+\bamu)\left[
P_1\left(2 \bam_1^2+2 L^2/J^2( P_{\tinf}-1) +m_0+\bamu\right)+
DQ_1\left(2 \bam_1^2+2 L^2/J^2 P_{\tinf} +m_0+\bamu\right)\right]}
{1+P_{\tinf}+DQ_{\tinf}}\right.
\\
\nn
&&\hspace*{9 cm}-\left.\frac{8\gamma DP_{\tinf}
\left(\bam_1-L/J P_{\tinf}\right)\left(1+\Kt_{\tinf}LQ_{\tinf}\right)}
{J\Kt_{\tinf}\bamu(1+DQ_{\tinf})^2}
\right\}
\EEA
\noindent where $Q_1$ is defined in (\ref{Q_1}) and
$P_1$ is the first order coefficient of $P$ of its expansion
in $\delta\mu_2$ (see (\ref{Pexp})).

To find the solutions to first order approximation in $\delta \mu_2(t)$
it is enough to keep in each of the (\ref{def:f1})-(\ref{def:g2})
only  terms  up to 
 $\cO(r \Upsilon)$ and $\cO\left(r\Upsilon/\left(\Gamma\mu_2 \right)\right)$.
Whether or not the terms of $\cO(r \Upsilon)$
 or those of   $\cO\left(r \Upsilon/\left(\Gamma^n\mu_2  \right) \right)$, 
with $n=1,2,...$ are the
most important depends
on the value of VFTH exponent $\gamma$: 
if $\gamma > 1$, $\cO\left(r \Gamma\mu_2\Upsilon\right)\ll \cO(r \Upsilon)$,  for
large times;
they are of the same order at $\gamma=1$,
while if $1/2<\gamma<1$, 
$\cO\left(r \Upsilon/\left(\Gamma\mu_2 \right)\right)\gg \cO(r \Upsilon)$.
Furthermore if $\gamma\leq 1/2$ also terms
 of order  $\cO\left(r \Upsilon/\left(\Gamma^2\mu_2 \right)\right)$ 
will be more  important or as
important as those of  $\cO(\Upsilon)$. For yet smaller values of $\gamma$
more and more terms of the kind $r \Upsilon/\left(\Gamma^n\mu_2 \right)$
will be  much  greater than  $\cO(r \Upsilon)$ in the aging regime.

As $t \to \infty$ the solutions (\ref{sec5:C11})-(\ref{sec5:C22}) coincides 
with the elements of the matrix (\ref{Hessian_inv}), the inverse of the 
Hessian of the free energy of the model, i. e. they coincides with the average 
squared fluctuations at equilibrium.

Once we have the equal time solutions we can solve the equations  (\ref{sec5:Cab})
for the two
times functions.

The function $\tilde{f}$ is now:
\BEA
{\tilde{f}}&\equiv& f_1+\frac{2\dot{m_1} g_1}{g_2-2 m_1 g_1}-g_1\frac{f_2-2m_1f_1}{g_2-2 m_1 g_1}=
\label{ft2}
\\ 
\nn
&=&-4\Upsilon(\Gamma-1)\frac{1+QD+P}{1+P}
-4\Upsilon\frac{\Gamma^2}{\mu_2} r
\gamma(m_0+\mu_2)\frac{QDP}{(1+P)^2} +\cO\left(\frac{\Upsilon \Gamma r}
{\mu_2}\right)
\EEA

The decoupled equations for $C_{11}(t,t')$ and $C_{12}(t,t')$ that we get are
always the (\ref{eq:C1b}). From them we can compute
 $C_{21}(t,t')$ and $C_{22}(t,t')$ as follows:
\BEA
&&C_{21}(t,t')\simeq 2\frac{
m_1(t)-P(t)L/J}{1+P}
C_{11}(t,t')
\\
&&C_{22}(t,t')\simeq \frac{2m_1(t)^2+2P(t)L^2/J^2+m_0+\mu_2(t)}{m_1(t)-P(t)L/J}
C_{12}(t,t')
\EEA

Using  the time evolution function $\tilde{h}$
for the considered  time-scale sector as
\BEQ
{\tilde{h}}(\tau)\equiv\exp\left(-\int_0^{\tau}{\tilde{f}}(t)dt\right)
\EEQ
\noindent  the solution of (\ref{eq:C1b}) comes out to be
\BEQ
C_{1b}(t,t')=C_{1b}(t',t')\frac{\tilde{h}(t')}{\tilde{h}(t)}+\cO(\mu_1\Upsilon)
\EEQ
In the leading terms of our expansion in $\delta\mu_2(t)$ and 
$\bamu$
the expressions 
for the $f$'s and $g$'s are given, for the case $T>T_0$ by 
(\ref{def:f1})-(\ref{def:g2}), 

Using (\ref{sec5:C11}), (\ref{sec5:C12}) and (\ref{ft2}),
 we get
\BEA
&&C_{11}(t,t')\simeq
\frac{1+P_{\tinf}}{1+Q_{\tinf}D+P_{\tinf}}
\left[m_0+\bamu
 +\cO(\delta\mu_2(t))\right]\exp\left\{-4\int_{t'}^{t}\Upsilon(t'')
(\Gamma(t'')-1)\frac{1+Q(t'')D+P(t'')}{1+P(t'')}\ dt''
\right\}\\
&&C_{12}(t,t')\simeq
\frac{2P_{\tinf}L/J-\bam_1}{1+Q_{\tinf}D+P_{\tinf}}
\left[m_0+\bamu+\cO(\delta\mu_2(t))
\right]\exp\left\{-4\int_{t'}^{t}\Upsilon(t'')
(\Gamma(t'')-1)\frac{1+Q(t'')D+P(t'')}{1+P(t'')}\ dt''
\right\}
\EEA

\subsection{Low temperature case: $T<T_0$}
Our approach allows also to study the regime 
below
below the Kauzmann temperature 
$T_0$.
In this last case, though, we have qualitatively
different behaviours depending on the 
value of $\gamma$, i.e. on the relative weight of $\mu_1$ and $\mu_2$.
We describe here the case $\gamma > 1$, where $\mu_1 \ll \mu_2$ (see 
(\ref{sol:mu11})).
For $\gamma>1$, according to the  results shown in section \ref{sec4},
it is, indeed,  not necessary to introduce any effective 
thermodynamic parameter other than the effective temperature,
and the analysis can be carried out in a way similar to the one  
of the previous case.
In expanding the time dependent coefficients of $C_{ab}$ in the 
equations of motion ($f_{1,2}$ and $g_{1,2}$ given in equations
(\ref{def:f1})-(\ref{def:g2}) ) we have now to take into account 
that $r$ never vanishes, while on the contrary the asymptotic value
of $\mu_2(t)$, denoted by $\bamu$, is zero.
The leading terms in the $f_k$ and $g_k$ ($k=0,1,2,3$)
 are in this case those of 
$\cO(\Upsilon\Gamma/\mu_2)$.
The sub-leading terms are those  of order   $\Gamma\Upsilon$ and 
$\Gamma\mu_2\Upsilon$ (coming always with $r$ as a multiplicative factor)
and those of order $\Gamma$. 
All these terms are diverging terms, in the limit $t \to \infty$,
and hence terms of $\cO(1)$ are now negligible with respect to them.
 They  would lead to corrections to the FDR of order $1/\Gamma\ll \mu_2$.

The equations of motions for the  equal time correlation functions are
identical to the (\ref{eq:C11tt})-(\ref{eq:C22tt}). What change are
 the time dependent coefficients $f_{1,2}$ and $g_{1,2}$ 
and $r=r_{\tinf}+\cO(\delta\mu_2(t))$, 
where $r_{\tinf}$ is defined in  (\ref{rinf}).

Solutions to these equations are obtained, as before, in the adiabatic
approximation
and expanding 
all the functions in powers of $\delta\mu_2(t)$ and $\bamu$:
\BEQ
C_{ab}(t)
= \left[2m_1(t)\right]^{a+b-2}\left\{\frac{1}{1+Q_{\tinf}D}
\left[m_0+\mu_2(t)\left(1-\frac{m_0Q_1D}{1+Q_{\tinf}D}\right)\right]
-\frac{m_0}{\Gamma}\left[
\frac{\alpha_1 r+\alpha_2 r^2}{(1+Q_{\tinf}D)^4}
+\delta_{a,2}\delta_{b,2}\gamma r \alpha_3
\right]
\right\}
+\cO(\mu_2^2(t)).
\label{sec5:Cabtt_2}
\EEQ
\noindent where $Q_1$ is the coefficient of the first order expansion of $Q$
given in (\ref{Qexp}), with $\bamu=0$ in this case and 
$\baw \to w_{\tinf}\equiv\sqrt{J^2 m_0+(D/\Kt_{eq})^2+T^2/4}$:
\BEQ
Q_1\equiv\frac{Q_{\tinf}}{(1+Q_{\tinf}D)m_0}
\left[\frac{J^2m_0(3 w_{\tinf}+T/2)}
{2 w_{\tinf}^2( w_{\tinf}+T/2)}-3P_{\tinf}\right]
\EEQ
\noindent and
\BEA
&&\alpha_1\equiv 1+3 DQ(1-2 P)+(DQ)^2(3-4P)+(DQ)^3 \ ,
\\
&&\alpha_2\equiv 4(1+DQ)\left[\Kt LQ(1+P)-DQ(1-P)\right]\ ,
\\
&&\alpha_3\equiv \frac{J m_1}{\Kt} Q\left(1+DQ-m_0\right)
\left(1+\Kt L Q\right)\left(5 -12 r+12r^2\right) \ .
\EEA

In the asymptotic limit this solutions do not coincide with 
(\ref{sec5:C11})-(\ref{sec5:C22}). 
That means that they are  also different from the static limit of
 the correlation functions
found in section \ref{sec2} from the inverse of the Hessian matrix 
(\ref{Hessian_inv}).
This is due to the fact that the static does not  take 
into account
the constraint (\ref{CONSTRAINT}) on the configuration space. 
Above the Kauzmann temperature the dynamics never reaches this constraint
so that, even if it is slowed down by its existence, it brings to the same static
results. But as soon as we perform the dynamics at $T_0$ or below it,
the asymptotic regime will never coincide with the equilibrium one. 
The system  will be stucked forever in one ergodic component of the phase
 space, artificially created by imposing the constraint in
the dynamics of the model, but not in the Hamiltonian.
The implementation of the constraint in the dynamics makes the variance 
$\Delta^2$ of the distribution of the random variables giving the updating
 of Monte Carlo dynamics (see (\ref{sec3:Delta})) diverging, when $T\leq T_0$.
The divergent factor of $\Delta^2$ is the quantity 
$\Gamma(t)$ appearing in the equations of motion so far discussed.
Going from a regime where the contribution of $\cO(1)$ are relevant ($T>T_0$) 
to another one where they are not even sub-leading, with respect
to $\cO(\Gamma)$,
we loose the static limit.

We find  the solutions of the equations  (\ref{sec5:Cab})
for the two-time correlation functions
following exactly the approach shown in the preceding section,
with the following expressions for the functions $\dot{m}_1$ and $\tilde{f}$:
\BEA
&&\dot{m}_1= 4\Gamma \mu_1\Upsilon +\cO(\mu_2^{\gamma}\Upsilon)
\\
&&
{\tilde{f}}=-4 \Upsilon \Gamma(1+QD)
-8\Upsilon\frac{QDP}{1+QD}(1-3r+2r^2)
+\cO(\mu_2\Upsilon)
\label{sec6:ftilde}
\EEA

The two-time correlation functions come out to be:
\BEA
&&C_{1b}(t,t')\simeq\frac{[2m_1(t')]^{b-1}}{1+Q_{\tinf}D}
\left[m_0+\mu_2(t')\left(1-\frac{m_0 Q_1 D}{1+Q_{\tinf}D}\right)
 +\cO(\mu_1(t'))\right]
\times
\\
 \nn
&&\hspace*{3 cm} \times\left\{
4\int_{t'}^t\left[1+Q(t'')D\right]\Upsilon(t'')
\Gamma(t'') dt''\right\}, \hspace*{1 cm} b= 1, 2.
\\
&&C_{2b}(t,t')\simeq 2 \left[m_1(t)\right]^{b-1}
C_{1b}(t,t') \hspace*{5 cm} b= 1, 2.
\EEA

For what concerns the response functions, from 
(\ref{sec5:G11_tt})-(\ref{sec5:G22_tt})  we get
\BEA
&&G_{11}(t,t^{\tp})=\frac{4\Upsilon\Gamma}{\Kt} 
 -\frac{2\Upsilon(1-2r)^2}{\Kt}
+\frac{8\Upsilon\left(\Gamma\mu_1\right)^2}{T_e}
\\
&&G_{12}(t,t^{\tp})=\frac{8m_1\Upsilon\Gamma}{\Kt} 
 -\frac{4m_1\Upsilon(1-2r)^2}{\Kt}
+\frac{16m_1\Upsilon\left(\Gamma\mu_1\right)^2}{T_e}
+\frac{16 r \Gamma\mu_1\Upsilon}{\Kt}
\\
&&G_{22}(t,t^{\tp})=\frac{16m_1^2\Upsilon\Gamma}{\Kt} 
 -\frac{8m_1^2\Upsilon(1-2r)^2}{\Kt}
+\frac{32m_1^2\Upsilon\left(\Gamma\mu_1\right)^2}{T_e}
+\frac{64m_1 r \Gamma\mu_1\Upsilon}{\Kt}
+\frac{8m_0(1-2r)^2\Upsilon}{\Kt}
\EEA
\noindent where this time the contributions $\Gamma\mu_1$ and 
$(\Gamma\mu_1)^2$ are both of order $\Upsilon$ and we take them into account.

The two-time behaviour of the response functions is as in (\ref{sec5:Gab}) 
with ${\tilde{f}}$ given by (\ref{sec6:ftilde}).

The last thing that we need, before computing the $T_e^{FD}$, is the derivative
\BEA
\partial_{t'}C_{11}(t,t')&=&\frac{\tilde{h}(t')}{\tilde{h}(t)}
\p_{t'}C_{11}(t',t')
-{\tilde{f}}(t')
\frac{\tilde{h}(t')}{\tilde{h}(t)}C_{11}(t,t')\simeq
\\ 
\nn 
&\simeq&
4 \Upsilon(t')
\frac{1}{1+Q_{\tinf}D}\left[m_0+\mu_2(t')\left(1-
\frac{m_0Q_1 D}{1+Q_{\tinf}D}\right)\right]
\left[1+Q(t')D\right]\Gamma(t') \times
\\ \nn &&\hspace*{5 cm}\times
\exp\left\{4\int_{t'}^{t}
\left[1+Q(t)D\right]\Upsilon{}(t'')\Gamma(t'') 
\ dt''
\right\}
\EEA

It follows
\BEQ
T_e^{FD}(t,t')\simeq T_e(t')\left[1+\cO\left(\frac{1}{\Gamma}\right)
+\cO(\mu_2^{1+\gamma})\right]=T_e^{FD}(t').
\EEQ
In this case $\cO(1/\Gamma)=\cO(\mu_1)$ is always smaller than $\cO(\mu_2)$ 
because 
$\gamma>1$: in the long time regime $T_e^{FD}(t)$ coincides with $T_e(t)$.

\subsection{Effective temperature from the  fluctuation formula}
A self consistent picture with an effective temperature should also imply 
that the same effective temperature  also governs other physical variables.
From the expression of $m_1(t;T)$ as function of $H$ we can compute
the quantity 
$\chi^{(fl)}\equiv\left.\frac{\partial m_1}{\partial H}\right|_{T,t}$
that is the contribution to susceptibility in a cooling-heating setup
to a change in the field $H$ at fixed time (also called {\em fluctuation susceptibility}).
In a cooling experiment the whole susceptibility can, indeed, be written as \cite{N00}
\BEA
\chi_{ab}\equiv \left.\frac{\p m_a}{\p H_b}\right|_T&=&
\left.\frac{\p m_a}{\p H_b}\right|_{T,T_e}+
\left.\frac{\p m_a}{\p T_e}\right|_{T,H_b}
\left.\frac{\p T_e}{\p H_b}\right|_{T}=\\
&=&\left.\frac{\p m_a}{\p H_b}\right|_{T,t}-
\left.\frac{\p m_a}{\p T_e}\right|_{T,H_b}
\left.\frac{\p T_e}{\p H_b}\right|_{T,t}+
\left.\frac{\p m_a}{\p T_e}\right|_{T,H_b}
\left.\frac{\p T_e}{\p H_b}\right|_{T}\equiv\\
&\equiv&
\chi^{fluct}_{ab}(t)\hspace*{3 mm}+\hspace*{7 mm}\chi^{loss}_{ab}(t)
\hspace*{7 mm}+\hspace*{7 mm}\chi^{conf}_{ab}(t)
\EEA

Here  we are considering an aging situation, so only the first term is relevant.
We can reasonably
assume that the $\chi^{fluct}_{ab}(t)$ can take the form
\BEQ
\chi^{fluct}_{ab}(t)=\left.\frac{\p m_a}{\p H_b}\right|_{T,t}=
N\frac{\left<\delta m_a(t)\delta m_b(t)\right>_{fast}}{T}+
N\frac{\left<\delta m_a(t)\delta m_b(t)\right>_{slow}}{T_e^{(fl)}(t)}.
\EEQ
\noindent where $\left<\ldots\right>_{fast/slow}$ is the average, 
respectively, over fast and slow processes. The fast ones
being governed by the heat-bath temperature,
the slow ones by some effective temperature $T_e^{(fl)}$
 depending on the time scale $t$.
Through $\chi^{fluct}_{ab}(t)$
  one can look at the  connection between the  fluctuation
effective temperature  $T_e^{(fl)}$, introduced in \cite{NPRL98},
and 
the other effective temperatures so far defined.
To work it out we start from:
\BEQ
\chi_{11}^{fluct}(t)\equiv\left.\frac{\p m_1}{\p H}\right|_{T,t}=
N\frac{\left<\delta m_1(t)\delta m_1(t)\right>}{T_e^{(fl)}}=
\frac{C_{11}(t,t)}{T_e^{(fl)}}.
\EEQ
\noindent Using the following expression of $m_1$ got from (\ref{sec3:eqm1}):
\BEQ
m_1(t;T,H)=-\frac{L}{J}+\frac{D}{J \Kt\left(m_1(t;T,H),m_2(t;T,H);T\right)}\ ,
\EEQ
\noindent the fluctuation susceptibility $\chi_{11}^{(fl)}$ turns out to be:
\BEQ
\left.\frac{\p m_1}{\p H}\right|_{T,t}=\frac{1}{\Kt(1+Q D)}+\cO\left(\mu_1\right).
\EEQ
\noindent Here we are neglecting terms like $\p \mu_1/\p H$ and 
 $\p \mu_2/\p H$, of order $\mu_1$ or higher (we deal with
the regimes [$T>T_0, \forall \gamma$]  and  [$T<T_0, \gamma>1$]
where $\mu_1\ll\mu_2$).
Taking the expressions 
(\ref{sec5:C11tt}) and (\ref{sec5:Cabtt_2})
we see that in both 
dynamic regimes 
that we are considering the leading term of $C_{11}$ can be written as
\BEQ
C_{11}(t,t)=\frac{m_0+\mu_2}{1+Q D}+\cO\left(\mu_2^{\gamma}\right)
\EEQ
\noindent and this leads to
\BEQ
T_e^{(fl)}=\Kt(m_0+\mu_2)+\cO(\mu_2^{\gamma}),
\EEQ
thus coinciding with (\ref{sec4:Te}) to the order of our interest, i. e. 
$\cO(\mu_2)$.
At higher orders there will be non-universalities.
If  $\gamma\leq 1$ the terms of $\cO(\mu_2^{\gamma})$ become 
dominant with respect to $\cO(\mu_2)$, leading
to the same situation that we had for $T_e^{FD}$ in (\ref{sec5:TeFD}),
namely the thermodynamic description does not lead to a unique
effective parameter.


 \renewcommand{\thesection}{\arabic{section}}
\section{Conclusions}
\setcounter{equation}{0}\setcounter{figure}{0}
\renewcommand{\thesection}{\arabic{section}.}

In this paper we consider a model that owns all the basic properties 
of a fragile glass,  built 
 by processes evolving on two  well separated time scales,
 representing the $\alpha$ and $\beta$ processes taking place in real
glassy materials.

Also, the model is provided with a constraint
applied to the harmonic oscillators dynamics, i. e. to the slow 
processes dynamics, in order to  reproduce the behaviour of a 
good glass former. 

Introducing a particular  Monte Carlo dynamics \cite{BPR,BPPR,N00}
 and developing it analytically, thus having the opportunity of probing 
it in more detail with respect to a numeric study,
 we found equations of motion that are in all respect 
 those typical of glass relaxation.

By means of the constrained dynamics we defined the Kauzmann temperature 
$T_0$ as the one at which  the constraint is reached, asymptotically, 
for the first time in a cooling experiment from high temperature.
There we showed how 
the real thermodynamic phase transition \cite{KAUZ}, taking place
 due to the breaking of the ergodicity in the  landscape of our model,
rich of degenerate minima, is characterized.

A detailed study of the dynamics was performed both above and below
 the Kauzmann temperature and for arbitrary values of
 the exponent $\gamma$ generalizing the typical 
VFTH behaviour, usually 
assumed for glasses, to
 $\tau_{eq}=\exp\left[A/(T-T_0)\right]^{\gamma}$. 
The dynamics in the aging regime
of both one time and two-time variables
has been carefully analyzed, including the corrections to this  regime, 
relevant at shorter times.

The decoupling of time scales is foundamental  for 
a generalization of equilibrium thermodynamics to systems 
far from equilibrium.
We tested on our exactly solvable model whether or not the 
 generalized approach holds,
 involving  one extra variable, namely the
{\em effective temperature}, in the
 description of  the non-equilibrium thermodynamics.
By effective temperature we mean a thermodynamic
quantity that would be  the temperature  of a system 
at equilibrium
visiting with the same frequency the same states 
that the real - out of equilibrium - system at temperature $T$
is visiting 
 on a given time-scale during its dynamics.
This kind of parameter appears in the thermodynamic functions
together with the heat-bath temperature and the fields coupled to
the system's observables and is coupled to the configurational entropy. 
In our work it has been 
first derived as the function of time (for given values of the heat bath
 temperature and of all the other parameters of the model) such that the
 evolving system out of equilibrium can be characterized by a probability
 measure
of the configurations having a Boltzmann-Gibbs form with a factor 
$1/T_e$ instead 
of $1/T$  in front of the Hamiltonian.

Generally speaking,  in order to recast the out of equilibrium dynamics,
the history of a system that is  far from equilibrium can be 
expressed by more  than one effective parameter.
This happens when more than one time-scale are involved in the dynamic 
evolution of a system. In those cases to every time-sector will
correspond an effective temperature \cite{CKL}.
Moreover the number of effective parameters needed to make such a translation
 into a thermodynamic view 
can, in principle, be equal to the number of relevant observables  considered
 in every time sector.
For  certain dynamic regimes, determined by the temperature and by
the VFTH exponent $\gamma$, however,
the effective parameters  pertaining to 
processes having the same time scale become
equal to each other in time, for large times.

As we saw in section \ref{sec4},
in the dynamic regimes reported in section
\ref{sec3}, when the variation $\mu_1(t)$ with respect to equilibrium 
of the variable $m_1$
is much greater than the distance from the constraint, $\mu_2(t)$,
 the effective temperature alone is
enough for a complete thermodynamic description of the dominant
physic phenomena (the effective field $H_e=H$), 
while in the regimes where $\mu_1(t)$ is no more 
negligible with respect to $\mu_2(t)$, 
 the effective field $H_e(t)$ is also needed to map the dynamics
on long time scales
into a thermodynamic frame.

From the time behaviour of the slow varying observables in the aging 
regime we found in section \ref{sec4}
a VFTH relaxation time dependence on temperature above the
 Kauzmann transition and we derived the Adam-Gibbs relation 
between the relaxation time and the configurational entropy,
that we can explicitly  compute for
our model.

We have been  also able to study the dynamics of the system 
quenched to a temperature below the Kauzmann temperature. 
At long, finite time $t$ we see that it is 
possible to introduce an instantaneous relaxation time depending on 
the heat bath temperature in a non-trivial way but 
expressible in terms of the effective temperature. What we got in this way
is actually a VFTH law
 where  the heat bath temperature
has been substituted by a time dependent effective temperature $T_e(t)$
and its  asymptotic value $\baTe$ takes the place of
the Kauzmann temperature $T_0$.
Such a relation for the time scale of the aging dynamics below $T_0$
could hold very well in more general systems.

At equilibrium the heat bath temperature enters many 
relations that can be rigorously proved
and connected to each other in
the frame of thermodynamics. Out of equilibrium we miss first principles
to start with in the generalization of such a construction.
We do not have any guaranty, for instance, 
that a given definition of effective temperature,
done generalizing a given equilibrium formula,
 would match any other definition coming from the generalization
of another equilibrium formula. At equilibrium the heat bath temperature
enters the Boltzmann-Gibbs measure,  the laws of thermodynamics,
the fluctuation-dissipation theorem, different Maxwell relations, etc.
However, out of equilibrium we have to check whether a single definition of 
effective temperature is compatible with any other. Since all
effective temperatures have  definite limits for long times,
we should verify that these limits are identical (which happens always)
and that the leading approach to these limits coincide (which happens
only for $\gamma>1$). 
If that works out we find a way to completely recast the long time domain
of the out of 
equilibrium dynamics into the language of thermodynamics, 
on a given time sector, of course, well separated from the other time sectors
of the glassy dynamics. This behaviour may occur if the aging is so slow that
the system has time enough to  clearly set out an effective temperature 
before going to a lower value of it.

With this aim 
we also rederived the effective temperature
 from the fluctuation dissipation ratio and from the 
fluctuation formula connecting the susceptibility with the
fluctuations of  the slow variables of the system, namely $\sum_i x_i$.
In section \ref{sec5} we showed that the effective temperature $T_e^{FD}$
 defined as the fluctuation dissipation ratio
tends to the effective temperature $T_e$ 
that we got by the quasi static approach  only if $\mu_2^{\gamma}$
is negligible with respect to $\mu_2$.
However, this is true, if $\gamma$ is greater than one. 
Otherwise the corrections of order $\mu_2^{\gamma}$ are no more sub-leading
and change significantly the time evolution of $T_e^{FD}$.
 Already for $\gamma=1$, $T_e^{FD}\to T_e$
only in the infinite time limit, i. e. for time scales  longer than those
of the considered aging regime.
Even above the Kauzmann temperature
the value of the exponent $\gamma$ discriminates between different regimes.
For $\gamma > 1$ an out of equilibrium thermodynamics can be built
in terms of the  single additional effective parameter $T_e$.
For $\gamma\leq 1$,  $T_e$ alone does not give consistent results
in the generalization of the equilibrium properties to the non equilibrium 
case. In those cases one also needs an effective field $H_e$.
However, no universal behaviour for the  $T_e$, $H_e$
combination has been found.

We have seen that in case $\gamma>1$ for temperatures above
the Kauzmann temperature both the statics and the
dynamics in the aging regime can be described 
by  Vogel-Fulcher-Tammann-Hesse laws, see eqs. (\ref{VFTHeq}), (\ref{VFTHag}). 
Notice that these laws are not identical, the static one diverging
more strongly, due to a larger prefactor of the divergent 
term in the exponent.
We  also noticed that in this situation $\gamma>1$ the
aging (well) below the Kauzmann temperature can be described in a 
form very analogous to the aging above it. For this reason it is
meaningful to compare experiments in those two aging regimes, 
and in particular, to test whether the decay of measurable quantities,
like the energy, the volume or the magnetization, has a 
common temporal law in the full aging regime.

Finally we checked the consistency of the widely mentioned 
thermodynamic  picture by writing the first and second law
using the effective temperature and we verified that it
 can be also computed from the generalization of the Maxwell relation 
at equilibrium giving the heat bath temperature as derivative of 
internal energy
with respect to the entropy (see (\ref{sec4:TeMax})), finding
 the same results we got from the other
derivations in the same validity limits. However, we also found that,
even within the solvable model of this paper, the applicability
of the thermodynamic picture may or may not be valid, 
depending on the question how fast the relaxation time diverges.

 \renewcommand{\thesection}{\arabic{section}}
\section*{Acknowledgments}
\setcounter{equation}{0}\setcounter{figure}{0}
\renewcommand{\thesection}{\arabic{section}.}
We thank F. Ritort for useful discussions.
The research of L. Leuzzi is supported by FOM (The Netherlands).

\renewcommand{\thesection}{\arabic{section}}
 \appendix{}
\section*{ Monte Carlo integrals}
\setcounter{equation}{0}\setcounter{figure}{0}
\renewcommand{\thesection}{\arabic{section}.}
\renewcommand{\theequation}{A.\arabic{equation}}
\label{app:A}
Here we present  the expressions of  the integrals that we use in computing 
the dynamics of the observables following the Monte Carlo method
explained in section \ref{sec3}. We recall that $x$, defined
in \ref{sec3:x}, is the energy difference bewteen the current configuration
of the system and the one proposed for the updating.
The variable $r$ (defined in (\ref{def:r}) )
is instead the distance of the effective temperature
$T_e$ from the heat-bath temperature (that is also 
the equilibrium value of $T_e$ in the dynamic regime above the Kauzmann
temperature).
First we define the abbreviation:
\BEQ
\Upsilon\equiv\frac{e^{-\Gamma}(1-r)}{\sqrt{\pi \Gamma}} \ ,
\EEQ
that is the leading term of the acceptance ratio of the Monte Carlo
dynamics given by:
\BEQ
\int dx W(\beta x)  p(x|m_1,m_2)=
\Upsilon\left(1-\frac{1-2r+4r^2}{\Gamma}+
+\frac{3}{4\Gamma^2}\left(1-4r+16r^2-24r^3+16r^4\right)+
\cO\left(\frac{1}{\Gamma^3}\right)\right).
\EEQ
Then we give the behaviour of the derivative  with respect to
time  of the energy
\BEQ
\int dx W(\beta x) x p(x|m_1,m_2)=
-4rT_e\Upsilon\left(1-\frac{3(1-2r+2r^2)}{\Gamma}+
\frac{15}{4\Gamma^2}\left(3-12r+28r^2-32r^3+16r^4\right)+
\cO\left(\frac{1}{\Gamma^2}\right)\right)
\EEQ
and of the variable $m_1$ (defined in (\ref{sec2:abbrev_m}))
\BEQ
\int  dx W(\beta x){\overline{y}}_1(x) p(x|m_1,m_2)=
4\mu_1\Upsilon\left(\Gamma-(1-3r+4r^2)+\cO\left(\frac{1}{\Gamma}\right)\right).
\EEQ

In section \ref{sec5} we compute the correlation and the response functions.
In order to find their time dependence we need the following derivatives. 
In these formulae we show the derivatives with respect to $m_1$ and
$m_2$, taken as indipendent variables, of the effective temperature $T_e$,
the variable $r$ and the leading term of the Monte Carlo acceptance ratio
$\Upsilon$.
They are:
\BEA
&&\frac{\p T_e}{\p m_1}=2 \Kt\left(P\frac{L}{J}- m_1\right),
\\
&&\frac{\p T_e}{\p m_2}= \Kt\left(P+1\right);
\EEA

\BEA
&&\frac{\p r}{\p m_1}=2\frac{1-3r+2r^2}{m_0+\mu_2}
\left(P\frac{L}{J}- m_1\right),
\\
&&\frac{\p r}{\p m_2}=\frac{1-3r+2r^2}{m_0+\mu_2}\left(P+1\right);
\EEA
and 
\BEA
&&\frac{\p \Upsilon}{\p m_1}=-\Upsilon\left[\frac{ m_1\gamma}{\mu_2}\left(2\Gamma+1\right)+2\frac{1-2r}{m_0+\mu_2} \left(P\frac{L}{J}- m_1\right)\right],
\\
&&\frac{\p \Upsilon}{\p m_2}=\Upsilon\left[\frac{ \gamma}{2\mu_2}\left(2\Gamma+1\right)-\frac{1-2r}{m_0+\mu_2} \left(P+1\right)\right].
\EEA

Furthermore,  we show the extensive computation of
 the coefficients of equations
(\ref{eq:C11tt})-(\ref{eq:C22tt}) and (\ref{sec5:Cab})
 for the dynamics of the two-time observables:
\BEA
f_0&\equiv&\partial_{m_1}\int dx  W(\beta x)  p(x|m_1,m_2)\simeq
- m_1\gamma\frac{\Upsilon}{\mu_2}\left[2\Gamma-1+4 r-8r^2+\frac{3}{2\Gamma}
\left(13-56r+136 r^2-160r^3+80r^4\right)\right]
\label{def:f0}
\\
\nn &+&\cO\left(\frac{\Upsilon}{\mu_2\Gamma^2}\right)
-2\frac{ \Upsilon}{m_0+\mu_2}\left(P\frac{L}{J}-m_1\right)\left[1-2r+
\frac{3-20r+40r^2-32r^3}{\Gamma}\right]+\cO\left(\frac{ \Upsilon}{\Gamma^2}\right)
\\
g_0&\equiv&\partial_{m_2}\int dx  W(\beta x)  p(x|m_1,m_2)\simeq
\gamma\frac{\Upsilon}{2\mu_2}\left[2\Gamma-1+4 r-8r^2+\frac{3}{2\Gamma}
\left(13-56r+136 r^2-160r^3+80r^4\right)\right]
\label{def:g0}
\\
\nn 
&+&\cO\left(\frac{\Upsilon}{\mu_2\Gamma^2}\right)
- \Upsilon\frac{1}{m_0+\mu_2}\left(P+1\right)\left[1-2r+
\frac{3-20r+40r^2-32r^3}{\Gamma}\right]+
\cO\left(\frac{\Upsilon}{\Gamma^2}\right)
\\
f_1&\equiv&\partial_{m_1}\int dx  W(\beta x) {\overline{y}}_1(x)  
p(x|m_1,m_2)\simeq
-4 m_1 \gamma \Upsilon \frac{\Gamma\mu_1}{\mu_2}\left[2\Gamma-3+6r-8r^2\right]+
\cO\left(\frac{ \Upsilon \mu_1 }{\mu_2}\right)
\label{def:f1}
\\ 
\nn 
&-&
16\Upsilon \frac{\Gamma\mu_1}{m_0+\mu_2}\left(\frac{L}{J}P-m_1\right)+
4\Upsilon\p_{m_1}\mu_1\left(\Gamma-1+3r-4r^2\right)+\cO(\mu_1\Upsilon),
\\
g_1&\equiv&\partial_{m_2}\int dx  W(\beta x) {\overline{y}}_1(x)
  p(x|m_1,m_2)\simeq
2 \gamma\Upsilon \frac{\Gamma\mu_1}{\mu_2}\left[2\Gamma-3+6r-8r^2\right]+
\cO\left(\frac{ \Upsilon \mu_1}{\mu_2}\right)
\label{def:g1}
\\ 
\nn 
&-&
8\Upsilon \frac{\Gamma\mu_1}{m_0+\mu_2}\left(P+1\right)+
4\Upsilon\p_{m_2}\mu_1\left(\Gamma-1+3r-4r^2\right) +\cO(\mu_1\Upsilon),
\\
f_2&\equiv&\partial_{m_1}\int dx  W(\beta x) {\overline{y}}_2(x)  p(x|m_1,m_2)
\simeq
2 m_1 f_1+\frac{2}{\Kt} f_3 +16 \Upsilon r \frac{L}{J}\frac{ P}{1+Q D}
 +\cO(\mu_1\Upsilon),
\label{def:f2}
\\
g_2&\equiv&\partial_{m_2}\int dx  W(\beta x) {\overline{y}}_2(x) 
 p(x|m_1,m_2)\simeq
2 m_1 g_1+\frac{2}{\Kt} g_3 +8\Upsilon r \frac{P}{1+QD} +\cO(\mu_1\Upsilon),
\label{def:g2}
\\f_3&\equiv&\partial_{m_2}\hspace*{-0.6mm}\int dx  W(\beta x)\  x \
 p(x|m_1,m_2)\simeq\hspace*{-0.6mm}
4m_1\gamma\Upsilon T_e\frac{r}{\mu_2}\left[\hspace*{-0.6mm}2\Gamma\hspace*{-0.6mm}
-\hspace*{-0.6mm}5\hspace*{-0.6mm}+\hspace*{-0.6mm}12r\hspace*{-0.6mm}
-\hspace*{-0.6mm}12r^2\hspace*{-0.6mm}-\hspace*{-0.6mm}
\frac{3}{\Gamma}\hspace*{-0.6mm}\left(\hspace*{-0.6mm}
3\hspace*{-0.6mm}+\hspace*{-0.6mm}9r\hspace*{-0.6mm}-\hspace*{-0.6mm}
54r^2\hspace*{-0.6mm}+\hspace*{-0.6mm}140r^3\hspace*{-0.6mm}-\hspace*{-0.6mm}
160r^4\hspace*{-0.6mm}+\hspace*{-0.6mm}80r^5\hspace*{-0.6mm}\right)\hspace*{-0.6mm}\right]\hspace*{-0.6mm}
\label{def:f3}\\
\nn
&+&\cO\left(\frac{\Upsilon r}{\mu_2\Gamma^2}\right)
-8\Upsilon\Kt \left(\frac{L}{J}P-m_1\right)
\left[1- 3r+4r^2-\frac{3}{\Gamma}\left(1-5r+12r^2-14r^3+8r^4\right)\right]+
\cO\left(\frac{\Upsilon}{\Gamma^2}\right)
\\
g_3&\equiv&\partial_{m_2}\hspace*{-0.6mm}\int dx  W(\beta x) x 
 p(x|m_1,m_2)\simeq\hspace*{-0.6mm}
-2\gamma\Upsilon T_e \frac{r}{\mu_2}\left[\hspace*{-0.6mm}2\Gamma\hspace*{-0.6mm}-
\hspace*{-0.6mm}5+\hspace*{-0.6mm}12r\hspace*{-0.6mm}-\hspace*{-0.6mm}12r^2
\hspace*{-0.6mm}-\hspace*{-0.6mm}
\frac{3}{\Gamma}\hspace*{-0.6mm}\left(\hspace*{-0.6mm}3+\hspace*{-0.6mm}
9r\hspace*{-0.6mm}-\hspace*{-0.6mm}54r^2\hspace*{-0.6mm}+\hspace*{-0.6mm}
140r^3\hspace*{-0.6mm}-\hspace*{-0.6mm}160r^4\hspace*{-0.6mm}+\hspace*{-0.6mm}
80r^5\hspace*{-0.6mm}\right)\hspace*{-0.6mm}\right]\hspace*{-0.6mm}
\label{def:g3}\\
\nn &+&\cO\left(\frac{\Upsilon r}{\mu_2\Gamma^2}\right)
-4\Upsilon\Kt \left(P+1\right)
\left[1- 3r+2r^2-\frac{3}{\Gamma}\left(1-5r+12r^2-14r^3+8r^4\right)\right]
+\cO\left(\frac{\Upsilon}{\Gamma^2}\right)
\\
\EEA
\noindent All partial derivatives with respect to $m_1$ have 
been computed keeping $m_2$ fixed and vice-versa. At this stage 
time has not yet been introduced.  Introducing it we are able
 to make an expansion of (\ref{def:f0})-(\ref{def:g3})
in powers of $\mu_2(t)$.
In the  formula we already performed such an expansion,
 breaking it at 
$\cO\left( \Upsilon/(\Gamma\mu_2 )\right)$ that is more than sufficiently
refined to derive the dynamics of the correlation and response functions
in all the regimes of our interest.


\addcontentsline{toc}{chapter}{}
\begin{thebibliography}{99}

\bibitem{ANGELL} C.A. Angell, Science {\bf 267} (1995) 1924.
\bibitem{NPRL98} Th.M. Nieuwenhuizen, Phys. Rev. Lett. {\bf 80} (1998) 5580.
\bibitem{N00} Th.M. Nieuwenhuizen, Phys. Rev. E, (2000)
\bibitem{BBM} A. Barrat, R. Burioni, M. Mezard, J. Phys. A: Math. Gen.
{\bf{29}} (1996)  1311.
\bibitem{BCKM} J. P. Bouchaud, L. Cugliandolo, J. Kurchan, M.Mezard, in 
{\em Spin Glasses and Random Fields}, A. P. Young, ed.  (World Scientific, 
Singapore, 1998), p. 161.
\bibitem{VFTH1}  H. Vogel,  Physik. Z. {\bf 22} (1921) 645.
\bibitem{VFTH2} G.S. Fulcher, J. Am. Ceram. Soc. {\bf 8} (1925) 339.
\bibitem{VFTH3} G. Tammann and G. Hesse, Z. Anorg. Allgem. Chem. {\bf 156}
 (1926) 245.
\bibitem{AG} G. Adam and J.H. Gibbs,
J. Chem. Phys. {\bf 43} (1965) 139.
\bibitem{ParisiVF} G. Parisi, 
in {\it The Oscar Klein Centenary}, U. Lindstr\"om ed.,
(World Scientific, Singapore, 1995); cond-mat/9411115.
\bibitem{KAUZ} W. Kauzmann, Chem. Rev. {\bf 43} (1948) 219.
\bibitem{MP}
Mezard M., Parisi G., J. Chem. Phys., {\bf{111}}  (1999) 1076
\bibitem{CPV1}
B. Coluzzi, G. Parisi, P. Verrocchio, J. Chem. Phys. {\bf{112}} (2000) 2933.
\bibitem{CPV2}
B. Coluzzi, G. Parisi, P. Verrocchio, Phys. Rew. Lett. {\bf{84}} (2000) 306.
\bibitem{GO}
W. G\"otze, Z. Phys. B {\bf{56}} (1984) 139.
\bibitem{CHS}
A. Crisanti, H. Horner, H. J. Sommers, Z. Phys. B {\bf{92}} (1993) 257.
\bibitem{KT}
T.R. Kirkpatrick, D. Thirumalai, Phys. Rew. Lett. {\bf{58}} (1987) 2091.
\bibitem{CS} A. Crisanti and H.J. Sommers, Z. Phys. {\bf B87} (1992) 341.
\bibitem{BPR} L.L. Bonilla, F.G. Padilla, and F. Ritort,
Physica A {\bf 250 }(1998) 315
\bibitem{CK93}
L. F. Cugliandolo, J. Kurchan, Phys. Rev. Lett. {\bf {71}} (1993) 173.
\bibitem{NCM} Th.M. Nieuwenhuizen, 
Solvable model for the standard folklore of the glassy state,
cond-mat/9911052. 
\bibitem{KA} W. Kob, H.C. Andersen, Phys. Rew. E {\bf{47}} (1993) 3281.
\bibitem{BPPR} L.L. Bonilla, F.G. Padilla, G. Parisi, F. Ritort, Phys. Rew. B 
{\bf 54} (1996) 4170.
\bibitem{ST96} M. Schulz, S. Trimper, Phys. Rew. B {\bf{53}} (1996) 8421.
\bibitem{KA2} W. Kob, H.C. Andersen, Phys. Rew. E {\bf{48}} (1993) 4364.
\bibitem{GLAUBER} R. J. Glauber, J. Math. Phys., {\bf{4}} (1963) 294.
\bibitem{MCKENNA} G.B. McKenna, in {\it Comprehensive Polymer Science
2: Polymer Properties}, C. Booth and C. Price, eds. 
(Pergamon, Oxford, 1989), pp 311.
\bibitem{N2} Th.M. Nieuwenhuizen, J. Phys. A {\bf 31} (1998) L201;\\
 Th.M. Nieuwenhuizen, Phys. Rev. Lett. {\bf 79} (1997) 1317.
\bibitem{SKT}
F. Sciortino, W. Kob, Tartaglia, Phys. Rew. Lett. {\bf{83}} (1999) 3214;\\
W. Kob, F. Sciortino, Tartaglia, Europhys. Lett. {\bf{49}} (1999) 590.
\bibitem{CR1}
A.Crisanti, F.Ritort, Physica A {\bf 280} (2000) 155;
\bibitem{FV}
S. Franz, M. A. Virasoro, J. Phys. A {\bf{33}} (200) 891.
\bibitem{CR2}
A.Crisanti, F.Ritort, Europhys. Lett. {\bf{51}} (2000) 147.
\bibitem{TOOL}
A. Q. Tool, J. Am. Ceram. Soc. {\bf{29}} (1946) 240. 
\bibitem{CKL}
L. Cugliandolo, J. Kurchan, P. Le Doussal, Phys. Rew. Lett. {\bf{76}} (1996) 2390.


\end{thebibliography}
\end{document}